\documentclass[
 reprint,=
 amsmath,amssymb,
 aps,unsortedaddress]{revtex4-1}
\usepackage{xcolor}
\usepackage{graphicx}
\usepackage{dcolumn}
\usepackage{bm}
\usepackage{amsmath}

\newcommand{\be}{\begin{equation}}
\newcommand{\ee}{\end{equation}}
\newcommand{\mac}{\mathcal}
\newcommand{\n}{\nonumber \\}

\begin{document}

\title{Symmetric Fracton Matter: Twisted and Enriched}

\author{Yizhi You}
\affiliation{Princeton Center for Theoretical Science, Princeton University, 
NJ, 08544, USA}

\author{Trithep Devakul}
\affiliation{Department of Physics, Princeton University, 
NJ, 08544, USA}

\author{F.~J. Burnell}
\affiliation{Department of Physics, University of Minnesota Twin Cities, 
MN, 55455, USA}

\author{S.~L. Sondhi}
\affiliation{Department of Physics, Princeton University, 
NJ 08544, USA}

\date{\today}
\begin{abstract}

In this paper, we explore the interplay between symmetry and fracton order, motivated by the analogous close relationship for
topologically ordered systems. 
Specifically, we consider models with 3D planar subsystem symmetry, and show  that these can realize subsystem symmetry protected topological phases with gapless boundary modes. Gauging the planar subsystem symmetry leads to a fracton order in which particles restricted to move along lines exhibit a new type of statistical interaction that is specific to the lattice geometry. 
We show that both the gapless boundary modes of the ungauged theory, and the statistical interactions after gauging, are naturally captured by a higher-rank version of Chern-Simons theory.
We also show that gauging only part of the subsystem symmetry can lead to symmetry-enriched fracton orders, with quasiparticles carrying fractional symmetry charge.

\end{abstract}

\maketitle

\section{Introduction}

One of the key paradigm shifts in modern condensed matter physics has been the appreciation of the many ways that topology affects our understanding of phases of matter. This new understanding has interacted in various interesting ways with the older paradigm of spontaneous symmetry breaking, which has been part and parcel of our understanding of phases of matter for much longer. Famously, it was found that non-interacting electron systems may harbor non-trivial topology in their band structures in classes delimited by symmetry, leading to what are now known as symmetry-protected topological (SPT) phases \cite{Fu2007-xo,bernevig2006quantum,Kane2005-ml,Moore2007-jk,Ryu2010-lj,Kitaev2009-ju,Roy2009-an}.  Analogs of these phases also exist in systems of interacting bosons \cite{Vishwanath2013-pb,Chen2011-kz,Chen2011-et,Chen2012-oa,lu2012theory,Senthil2015-tp,wang2014interacting,bi2014anyon,you2014symmetry} and fermions \cite{fidkowski2011topological,wang2014classification,fidkowski2011topological,fidkowski2010effects,qi2017folding}.  Strongly interacting many-body systems may exhibit topological order\cite{wen1990topological,devakul2017mathbb,Bravyi2011-fl,Gaiotto2016-ba,levin2005string,hansson2004superconductors}, in which emergent quasi-particles have long-ranged statistical interactions that differ from those of the microscopic constituent fermions and bosons without the need to impose a symmetry requirement.  Finally, topologically ordered systems {\it can} come equipped with global symmetries and exhibit so-called symmetry enriched topological, or SET, phases which are more than the sum of their parts, since these anyonic quasi-particles naturally carry fractional symmetry charges\cite{Vishwanath2013-pb,Fu2007-xo,fidkowski2011topological,Pollmann2012-lv,Gu2014-lj,Chen2013-gq,wang2014interacting,Chen2012-oa,chen2011two,Ryu2010-lj,Cheng2015-ul,Roy2009-an}.


We now know that topologically ordered, SPT, and SET phases are intimately mathematically related.  In general, gauging a discrete global symmetry leads to topological order; the statistical interactions of this topological order diagnose whether the ungauged theory is an SPT \cite{Levin2012-dv,levin2012classification,bi2014anyon,Vishwanath2013-pb}.  Similarly the statistical interactions of the topological order tightly constrain  the possible fractional symmetry charges in an SET phase \cite{wang2017anomaly,barkeshli2014symmetry,metlitski2013bosonic,ryu2015interacting}. Finally, gauging only part of the global symmetry of an SPT generically yields an SET phase with non-trivial symmetry fractionalization.  

Recently, a new type of order in 3 dimensions---nicknamed fracton order---which has precursors in the study of glassiness\cite{Chamon2005-fc}, spin liquids\cite{yoshida2013exotic,Yoshida2013-of}, and quantum error correcting codes\cite{Haah2011-ny}, has drawn increasing attention\cite{Vijay2016-dr,Devakul2017-gg,Slagle2017-ne,Ma2017-qq,Halasz2017-ov,Hsieh2017-sc,Vijay2017-ey,Slagle2017-gk,Slagle2017-st,Williamson2016-lv,Ma2017-cb,Shi2017-bs}.  Fracton order is qualitatively different from topological order in 3 dimensions, most obviously because it has a subextensive ground state degeneracy that depends not only on the topology of the spatial lattice, but also on its geometry. Further, it is characterized by excitations that move on a manifold of dimension strictly lower than that of the lattice they live on. Finally, unlike for topological order, the appropriate field theoretic description for fracton orders is not yet completely understood, though considerable progress has been made in connecting it to both continuum \cite{Slagle2017-la,shirley2017fracton,cho2015condensation,Pretko2017-ej,gromov2017fractional} and discrete higher-rank gauge theories\cite{ma2018fracton,prem2017emergent,bulmash2018higgs,pretko2017subdimensional,ma2018higher,pretko2017generalized,pretko2017finite}.
Nevertheless, fracton order shares many of the defining features of topological order, including a strongly correlated liquid ground state with no long-ranged order parameter, and emergent point-like quasiparticles which, by virtue of their restricted mobility, can have non-trivial statistical interactions even in $D=3$.

This then leads to the question of how fracton order interacts with symmetries---are their analogs of SPT and SET phases? To answer the first question, a useful observation is that for fracton orders, the appropriate symmetry is not a global symmetry, but a so-called {\it subsystem symmetry}, which acts simultaneously on all sites in a given subsystem of the lattice.   
Unlike discrete global symmetries, which upon gauging yield topological order, gauging an appropriate subsystem symmetry yields a fracton order.  The subsystem can be either a 
plane (in which case the resulting fracton order is said to be Type I) or a fractal subset (leading to Type II fracton order)\cite{Vijay2015-jj,kubica2018ungauging,Vijay2016-dr,williamson2016fractal,yoshida2013exotic}. 
This relationship between fracton order and subsystem symmetry was first established for Ising symmetry, with each subsystem a conventional Ising paramagnet \cite{Vijay2016-dr}, and later generalized to include planar subsystem symmetries with more general discrete symmetry groups \cite{ma2018fracton,prem2017emergent,bulmash2018higgs}.

In recent work, we showed that subsystem symmetries acting on lines\cite{2018arXiv180302369Y} or fractals\cite{devakul2018fractal} can protect new non-trivial subsystem-SPT (or SSPT) phases. These are characterized by a ground state degeneracy that grows with the system's perimeter, which cannot be lifted without breaking the {\it subsystem} symmetry.  (The global symmetry alone, in contrast, is not sufficient to protect this boundary degeneracy).  

Here, we will explore the possibility of planar SSPT phases, which raise a fresh set of questions. For topological orders, it is well-established that if a phase with global symmetry is a non-trivial SPT, then the resulting gauged theory necessarily has a different topological order than that obtained from the conventional (non-SPT) paramagnet---it is said to exhibit a twisted version of the topological order. Are there twisted fracton phases? And to complete the mapping from topological phases, are there fracton SETs?


Motivated by the above considerations, the present work seeks to address the following inter-related questions.  (i) Can subsystem symmetries acting on planes (d=2) in 3 spatial dimensions (D=3) be non-trivial SSPT phases, with boundary modes that cannot be gapped without breaking subsystem symmetry?  (ii) If so, what impact does going from trivial subsystem-symmetric to SSPT phases have on the fracton order of the gauged theory?  
(iii) Are there fracton analogs of SET phases, in which fractons carry fractional symmetry charges?  The analogies that underlie (i) and (ii) are displayed in Fig.~1.

Our main results are the following.  First, we present a family of models that realize planar (i.e. subsystem dimension 2, which we refer to as $d=2$) SSPT phases with Ising ($Z_2$) and $Z_2 \times Z_2$ symmetry.  We show that upon gauging, these SSPT phases lead to fracton orders with qualitatively different statistical interactions than their non-SSPT counterparts, and we clarify the nature of the new type of statistical interaction. Using a method similar to that of Ref. \cite{Levin2012-dv}, we relate these new statistical interactions to the existence of ungappable boundaries.  We then present a field theoretic description of these phases -- a higher-rank version of Chern-Simons theory -- that captures these gapless boundaries.  This field theory has the interesting feature that it can seemingly describe phases that are not SPT, in the sense that the boundary modes are topologically protected (as for Chern-Simons theory), rather than symmetry protected.  Finally, we describe two models with both fracton order and symmetry fractionalization (the fracton analogue of SET phases).   These harbor quasiparticles that are either immobile, or move on lines, with $\mathcal{T}^2 = -1$.   In the process, we develop a general picture of how the fracton ground state can be decorated with objects charged under the global symmetry to yield an SET phase.


In more detail, our discussion proceeds as follows.  Section \ref{XCubeReview} reviews the relevant background information about planar subsystem symmetry, and its relationship to fracton order upon gauging.  
In section \ref{tp}, we present an exactly solvable $3D$ model similar in spirit to the $2D$ plaquette Ising model\cite{Levin2012-dv},  with a ``twisted" $Z_2$ subsystem symmetry.  We show that gauging this twisted $Z_2$ subsystem symmetry leads to a twisted fracton order, with a ``lineon" (or excitation confined to move only in one dimension)  with an analogue of semionic self-statistics. We also show that a variant of the argument of Ref.\cite{Levin2012-dv} can be used to show that these non-trivial statistics imply the presence of symmetry-protected gapless boundaries. In section \ref{tptp} we discuss how this extends to $Z_2 \times Z_2$ symmetry, which has the new feature that lineons may harbor both non-trivial self and non-trivial mutual statistics. 

In section \ref{cs}, we use these results to motivate  a higher-rank variant of Chern-Simons theory, which we show has both gapless boundary modes and 
a form of dipole Hall response in the presence of a (rank-2) electric field. 

Finally, in sections \ref{set} and \ref{FractSet}, we introduce models with $Z_2 \times \mathcal{T}$ subsystem symmetry, whose corresponding gauge theory is a symmetry enriched fracton phase whose lineon (or fracton, in Sec. \ref{FractSet}) excitation has $\mathcal{T}^2 = -1$, which for bosonic systems should be interpreted as a fractional symmetry charge under time reversal. These can be viewed as subdimensional spin liquids, where the system supports deconfined spinon excitations with restricted mobility.
Our construction proceeds by a decoration procedure that may be extendable to other symmetry groups.

\begin{figure}[h]
  \centering
      \includegraphics[width=0.5\textwidth]{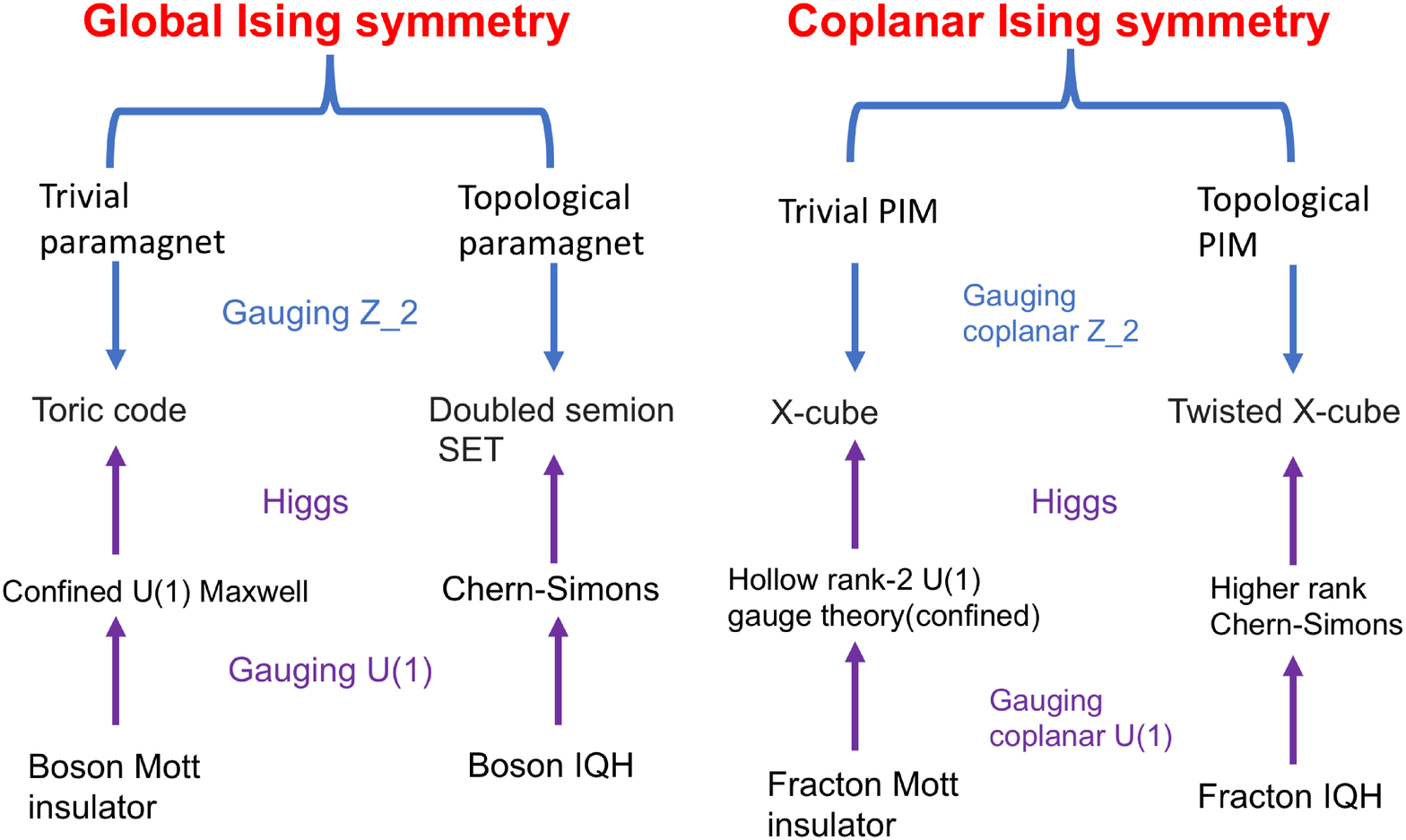}
  \caption{Comparison of the relationship between subsystem SPT phase, twsited fracton theory and higher rank Chern-Simons term with their respective counterparts in topologically ordered systems.  Here the Higgs phases are condensates of objects with charge $2$, leading to the nontrivial orders.} 
  \label{overall}
\end{figure}


\section{Review of planar subsystem symmetries and fracton models}
\label{XCubeReview}

To set the stage, in this section we will review the plaquette Ising model (PIM)\cite{Johnston2016-nf}, which is the simplest example exhibiting planar subsystem symmetry in 3D.  We utilize a ``domain frame"~\cite{huang2018cage} description of the resulting ground state, which we will use extensively in later sections.  We also review the canonical procedure to gauge the planar subsystem symmetry, and the basic properties of the resulting fracton topological order\cite{Vijay2015-jj,Vijay2016-dr,shirley2017fracton,Slagle2017-gk,2018arXiv180506899S}.  

\subsection{Plaquette Ising model in 3D: Hamiltonian and ground states}
The 3D plaquette Ising model consists of Ising spins on the sites of a cubic lattice, with the Hamiltonian:
\begin{align} 
&H(k)=- J \sum_P \prod_{i \in P} S^z_i  -h\sum_i S^x_i
\label{pising}
\end{align}
The first term is a quartic interaction between the four spins on the same plaquette $P$, while the second term is the external transverse field. 

The Hamiltonian (\ref{pising}) commutes with any process that flips an even number of spins on each plaquette -- i.e. with any operation that flips all spins in a certain plane.  Hence unlike the conventional Ising model, which has a global $Z_2$ symmetry, the PIM has $L_x+L_y+L_z-2$  independent $Z_2$ symmetry operations (one for each plane, and the $-2$ comes from the fact that flipping all $xy$ planes is the same as flipping all $yz$ or $zx$ planes).
This vastly enlarged symmetry group is known as $Z_2$ subsystem symmetry ($Z^{sub}_2$). 

\begin{figure}[h]
  \centering
      \includegraphics[width=0.25\textwidth]{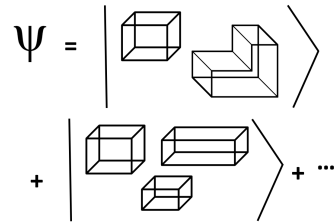}
  \caption{Domain frame condensate for the plaquette Ising model in the paramagnetic phase.} 
  \label{trivialwf}
\end{figure}

We will primarily be interested in the paramagnetic phase occurring for $h \gg J$.  This leads to a symmetric ground state, with all spins polarized along the $x$-axis.  For our purposes it is convenient to express the wave function in terms of the eigenvalues of the first term in Eq. (\ref{pising}).  We represent the spin configuration by drawing a line through each paquette where  $ S^z_i S^z_j S^z_k S^z_l  =-1$.   
These lines cannot terminate, and cannot form isolated closed loops.  Instead, they must form the domain frame structure like those shown in Fig~\ref{trivialwf}.
When the domain frames have proliferated, the system is in its paramagnetic phase and preserves the $Z^{sub}_2$ symmetry.

We note that since the subsystem symmetry commutes with $H$, it does not change the domain frame configuration.  Thus each domain frame represents an extensive number of actual spin configurations, related by the $L_x + L_y + L_z-2 $  $Z_2$ subsystem symmetries.  (This is markedly different than a domain wall description, which can represent only two distinct spin configurations).  The transverse field in Eq. (\ref{pising}) ensures that the ground state is a superposition of all possible spin configurations for each domain frame, with equal coefficients.  In general, we will take a single domain frame to represent this particular superposition over spin configurations.


\subsection{Gauging the Plaquette Ising Model: fracton topological order} \label{Sec:GaugeReview}

Part of the interest in the PIM stems from the phase obtained by gauging the $Z^{sub}_2$ symmetry. 
We will utilize a generalized gauging procedure~\cite{Vijay2016-dr,Williamson2016-lv} which, when applied to the PIM,
results in the X-cube model exhibiting $3D$ fracton order.  
Here we will review how this gauging process is carried out, as well as the key features of the resulting fracton phase.

The procedure for gauging the subsystem symmetry is similar in spirit to that used to gauge the global $Z_2$ symmetry of the ordinary Ising model.
Following Ref. \cite{Vijay2016-dr,Williamson2016-lv}, to gauge the symmetry we first add a new Ising variable $\sigma^z_P$ to the center of each plaquette $P$, and couple it to the four plaquette spins:  
\begin{align}  \label{HPIMGauge}
&H=-\sum_P  J \sigma^z_{P}\prod_{i \in P}S^z_i   -h \sum_{i} S^x_i
\end{align}
 This ensures that we may flip individual spins without incurring an energy cost, provided we also flip an appropriate set of $\sigma^z$.   
 Second, we enforce the constraint at each site $i$,
 \begin{align} \label{GaugeConstr}
&\prod_{P | i \in P}\sigma^x_{P} = S^x_i
\end{align}
 where the product runs over the 12 plaquettes touching this site.  
This is the analogue of Gauss' law in this gauge theory\cite{Vijay2016-dr,bulmash2018higgs,ma2018fracton,pretko2017fracton}.  

\begin{figure}[h]
  \centering
     \includegraphics[width=0.25\textwidth]{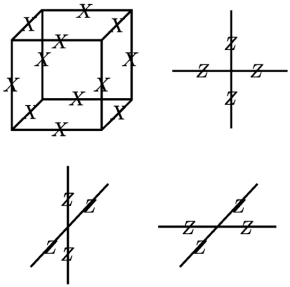}
  \caption{Couplings in X-cube model. The 12 spin interaction on the cube indicates the Gauss law constraint. The four spin in the vertex in the $\alpha-\beta$ plane describes the gauge fluctuation.} 
  \label{xcube}
\end{figure}

Quite generally, we can use this Gauss law to reduce the variables to those of the gauge field alone. Deep in the paramagnetic phase ($h\gg J$), the effective Hamiltonian for the gauge fields is especially simple. It is derived, by using Eq.~(\ref{GaugeConstr}) to replace $S^x_i$ with $\prod\sigma^x$, and then keeping only those products of the plaquette terms in Eq. (\ref{HPIMGauge}) that commute with this product. 
Since the gauge fields live on plaquette centers, it is convenient to depict these interactions on the dual lattice, whose links penetrate the plaquette of the original cubic lattice (Fig. \ref{xcube}). The dual lattice is also a cubic lattice, with a gauge variable $\sigma$ living on each link.

Thus after gauging, and projecting to the low-energy Hilbert space for $h \gg J$, and setting the resulting couplings to 1 (which does not alter the resulting physics), we obtain an exactly solvable commuting projector Hamiltonian, known as the X-cube model \cite{Vijay2016-dr}:
\begin{align}   \label{HXCube}
&H_{XC}=-\sum_v \sum_{(\alpha\beta)\in\{ab,bc,ca\}} \prod_{i\in C_v^{\alpha\beta}} \sigma^z_i - \sum_{Cube} \prod_{i\in Cube}  \sigma^x_i 
\end{align}
where $a,b,c$ refer to the three principal axes of the cubic lattice,
and $i$ labels a link on this dual lattice.
$C_v^{\alpha\beta}$ is the set of four edges that lie in the $\alpha-\beta$ plane, and ending at vertex $v$, as shown in Fig.~[\ref{xcube}]. 
We will call this set of edges a ``cross"  in the $\alpha-\beta$ plane. The second term is simply the left-hand side of Eq. (\ref{GaugeConstr}), now written on the dual lattice; here $\prod_{i\in Cube}$ includes $\sigma$ spins on the 12 links of the cube. 

\begin{figure}[h]
\centering
 \includegraphics[width=0.45\textwidth]{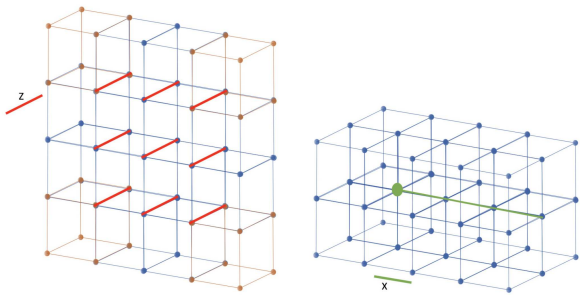}
\caption{L: The charge excitation generated by the 2d membrane(red links) operator.  At each corner of the membrane, there is a cube where $\prod_{i\in c}  \sigma^x_i =-1$, which contains a charge (fracton) excitation;
R: The lineon(flux) excitation generated by a straight string(green). The lineon excitation lives at the end of string.}
\label{dipole}
\end{figure}

The X-cube model is a canonical example of a model exhibiting Type-I fracton order\cite{Vijay2016-dr}.  To understand what this means, let us examine the low-lying excitations of $H_{XC}$.  
When $\prod_{i\in \text{Cube}}  \sigma^x_i =-1$, the constraint (\ref{GaugeConstr}) dictates that a $Z_2$ charge excitation, with $S^x=-1$, lies in the center of the cube.  
From Fig.~[\ref{xcube}], we see that these charge excitations must be created in multiples of 4. 
Further, they are not independently mobile:
A quadruplet of well-separated charges is created by an operator $\prod_{i \in \text{Membrane}}  \sigma^z_i$, with a {\em square} membrane, cutting through links as shown in Fig.~[\ref{dipole}].   
A pair of $Z_2$ charges can be viewed as a dipole excitation\cite{Pretko2017-ej,pretko2017fracton}, which can move in the $2d$ plane perpendicular to the dipole moment. 
Particles with the property that they are individually immobile, but may be moved in conjunction with other particles, are called {\it fractons}.

The second type of excitation, which is associated with flux of the $\sigma$ gauge field, lives at vertices where $\prod_{i\in C_v^{\alpha\beta}} \sigma^z_i = -1$.   These fluxes are pair-created by the line operator $\prod_{i \in \text{Line}}  \sigma^x_i$ in Fig.~[\ref{dipole}].  If the line is in the $\alpha$ direction, then at each end point we have $\prod_{i\in C_v^{\alpha\beta}} \sigma^z_i=\prod_{i\in C_v^{\alpha\gamma}} \sigma^z_i=-1$, for $\alpha,\beta,\gamma$ all different directions.  Notice that the line must be straight: if it changes direction, additional  excitations are created at the corners. Thus there are three types of flux excitations, each of which can move along a particular direction.   We will call excitations with this property lineons. 

The fracton order is defined by the sub-dimensional nature of these two excitations, together with their nontrivial mutual statistics. From the form of the line and membrane operators it is easy to verify that the lineon and fracton have a nontrivial statistical interaction\cite{Ma2017-cb,Ma2017-qq,Slagle2017-gk,Devakul:2018aa}, resulting in a $\pi$ statistical phase if a pair of fractons crosses a lineon.  As a consequence, it can be shown that the theory has size dependent topological degeneracy when placing on three torus\cite{Haah2011-ny,He2017-eq}.

\section{Topological plaquette paramagnet and twisted fracton order}\label{tp}

In the search of 2D SPT states with $Z_2$ symmetry, 
Levin and Gu\cite{Levin2012-dv} showed that the ordinary Ising paramagnet can be modified to obtain a topological Ising paramagnet -- a distinct phase with the same unbroken symmetry, and protected gapless edge modes.  Upon gauging, this second model yields a ``twisted" version of the Ising gauge theory, in which the point-like excitations have different self- and mutual- statistics.

In this section we present a variant of the 3D plaquette Ising model that has been modified in a  similar spirit.  We will show that, upon gauging, this model realizes the fracton topological order first introduced by Ref.\cite{Ma2017-qq}, which is essentially a twisted variant of the X-cube model in which the fracton and lineon excitations have different self- and mutual- statistics.  To clarify this, we will discuss a new type of statistical interaction, which we dub ``boxing", associated with non-trivial lineon self-statistics. Finally we will leverage insight from these statistical interactions to show that, like the topological Ising paramagnet, this system has symmetry-protected gapless boundary modes.

\subsection{Topological plaquette paramagnet in 3$D$}


The intuitive idea of Levin and Gu's construction\cite{Levin2012-dv} is to modify the Hamiltonian such that in the ground state, configurations with even and odd numbers of domain walls appear with a relative minus sign.  For $d=2$ subsystem symmetries, the analogue of a domain wall is a domain frame.  Indeed, our model can be viewed as a modified plaquette Ising model, in which configurations with even and odd number of domain walls on the frames appear with a relative minus sign.

\begin{figure}[h]
  \centering
      \includegraphics[width=0.2\textwidth]{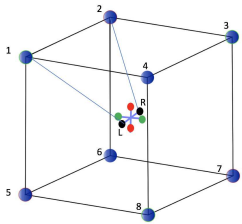}
  \caption{The spin model on the BCC lattice. Each corner of the cube contains an Ising spin $S_0$. The cube center contains 3 spin dipoles $(S_a,S_b,S_c)$. The spin interaction appears between four spin $S_0$ on the same cube face, as well as two $S_0$ together with the spin dipole $(S_\alpha)$ on the triangle.} 
  \label{aaa}
\end{figure}

To achieve this sign structure, however, it is advantageous to work on a more complicated lattice.  Here we work on a cubic lattice with 7 spins per unit cell, arranged as shown in Fig.~[\ref{aaa}].  On the corners of each cube, there is a single Ising spin $S_0$. In the center of the cube, there are 3 Ising spins, $S_a,S_b,S_c$, which we will collectively refer to as spin dipoles due to the nature of their couplings with the remaining spins. 
Each spin dipole is associated with a particular cubic axis: The red spins $(S_a)$ with the $a$ direction, the green spins  $(S_b)$ with $b$, and the black spins $(S_c)$ with $c$. 

\begin{figure}[h]
  \centering
      \includegraphics[width=0.4\textwidth]{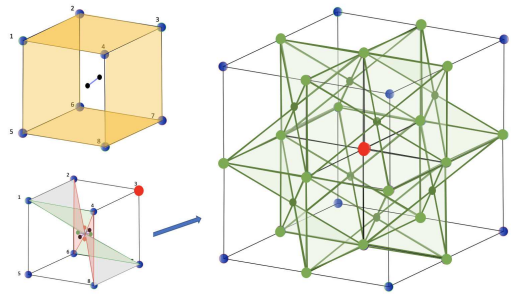}
  \caption{Left top: The spin dipole $S^{x}_\alpha$ oriented along the $\alpha$-axis ($\alpha = a,b,c)$ in the cube's center interacts with  the four plaquettes on the side face (yellow) of the cube parallel to the $\alpha$ direction. Left bottom and Right: The spin $S^x_0$ on the site (red) interacts with the nearby 48 triangles (green) that form a closed Hexoctahedron surface enclosing $S_0$. In the left bottom figure, we plot the 6 triangle at the (111) corner. The are 8 corners and each contains 6 triangles with a total of 48.} 
  \label{ddd}
\end{figure}

The Hamiltonian of the topological plaquette paramagnet is :
\begin{align} 
&H_0=- \sum_i \left ( S_{0,i}^x i^{F_0} -S^{x}_{a,i} i^{F_{a,i}} -S^{x}_{b,i} i^{F_{b,i}} -S^{x}_{c,i} i^{F_{c,i}} \right )\nonumber\\
&F_{\alpha\neq 0} =\sum_{P_\alpha} (1-\prod_{i\in P_\alpha} S^z_{0,i} )/2\nonumber\\
&F_0=  \sum_{\substack{T_\alpha\\ (ijk\in T_\alpha)}}  (1-S^z_{0,i} S^z_{0,j} S^{z}_{\alpha,k} )/2 
\label{abc}
\end{align}
$H_0$ contains a transverse field for each spin. 
However,  each transverse field is decorated with a special sign structure $i^{F}$, where $F$ depends on the local domain frame configuration.  
For example $F_a$, which dictates this sign structure for  $S^{x}_a$, the spin associated with the $a$ direction, counts the number of domain frame lines passing through the four neighbouring plaquettes $P_a$ that are parallel to $\hat{a}$ (see Fig.~[\ref{ddd}]).  $F_b$ and $F_c$ are defined analogously.  
Meanwhile $F_0$, which determines the sign structure for a corner spin $S_0$, counts the number of domain frame lines crossing the 48 triangles of the Hexoctahedron surface enclosing $S_0$ (see Fig.~[\ref{ddd}]).  
Here $T_a$ denotes a triangle containing the (black) spin associated with  
the $\hat{a}$ direction, and two neighbouring corner spins separated in the $\hat{a}$ direction; an example is shown in Fig.~[\ref{aaa}].  $T_b$ and $T_c$ are defined analogously.  
We denote the 3 sites involved in each triangle $ijk\in T_\alpha$ such that $i,j$ refer to $S_0$ spins, while $k$ refers to the $S_\alpha$ dipole.
The sum is over $T_\alpha$ is over 48 such triangles.
This geometry is complicated, but will become clearer as we transition on to the dual lattice. 

Rest assured that despite the $i^F$ factors, the Hamiltonian is Hermitian.  
This is because the number of plaquettes for which $\prod S^z =-1 $ is {\it even} on any non-contractible surface-- and in particular, on those used to construct our $F$ terms. Thus each $F$ is necessarily even,  
and in practice the phases are all real.

As the $F$ terms in the Hamiltonian involve products of spins on the same plaquette/triangle,  the system is invariant under 
$Z^{sub}_2$ subsystem symmetry operations of the form: 
\begin{align} 
Z^{sub}_2: & S^z_0(r_a=a_0) \rightarrow -S^z_0(r_a=a_0),\nonumber\\
& S^{z}_a(r_a=a_0+\frac{1}{2}) \rightarrow -S^{z}_a(r_a=a_0+\frac{1}{2}) \nonumber\\
& S^{z}_a(r_a=a_0-\frac{1}{2}) \rightarrow -S^{z}_a(r_a=a_0- \frac{1}{2}) 
\end{align}
where $\vec{r}=(r_a,r_b,r_c)$ is the position vector of the spin $S_0$,
and similarly for planes orthogonal to the $b$ and $c$ axes.
Each $Z^{sub}_2$ symmetry acts on a 
plane (say perpendicular to the $\hat{a}$ direction) and flips all $S^z_0$ in that plane, as well as all the dipole spins associated with the vector normal to the plane (i.e. $\hat{a}$) both above and below the plane. 

Although the Hamiltonian has a complicated, indeed horrifying, form the intrinsic nature of our model is transparent. 
First, note that all of the terms commute.  This can be checked by straightforward (albeit tedious) algebra; the key is that  $S^x_i$ changes the value of each $F$ that includes site $i$,  and that if $F_j$ includes site $i$, then also $F_i$ includes site $j$. 

\begin{figure}[h]
  \centering
      \includegraphics[width=0.35\textwidth]{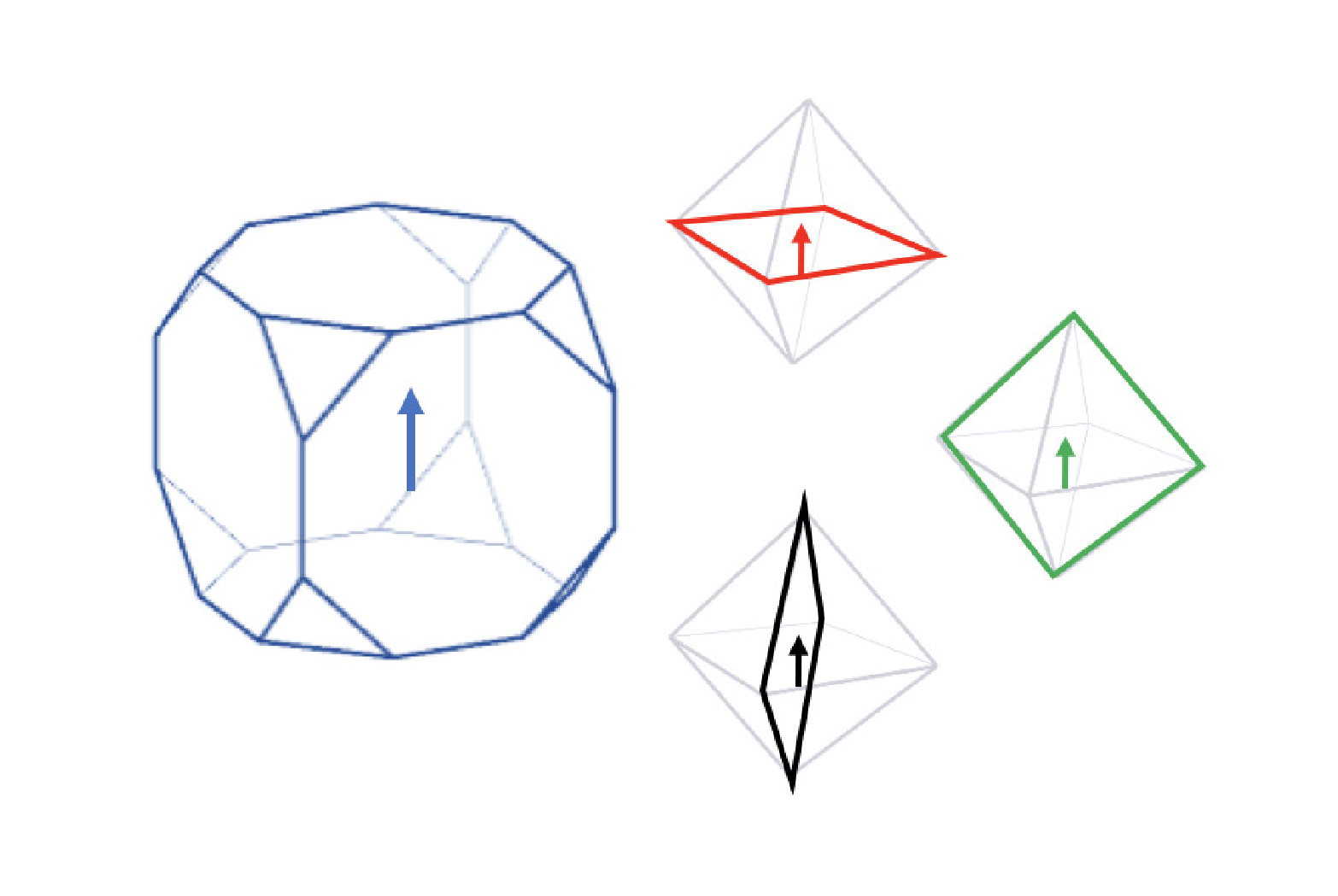}
  \caption{The domain frame configuration for $S^z_0$ and the spin dipoles $S^z_i$. Flipping the spin for $S^z_0$ creates a domain  frame on a truncated cube (left). Flipping a spin dipole for $S^z_i$ creates an domain plaquette in the direction perpendicular to the dipole (right). The three orthogonal domain plaquettes form an octahedron living on the corner between truncated cubes.} 
  \label{frame}
\end{figure}

Because all terms in $H_0$ commute, we can understand the ground state by examining the individual terms.  
First, away from a boundary there is one term for each spin, and the system has a unique paramagnetic ground state.
On the dual lattice, flipping one $S^z_0$ results in the domain frame defect along the truncated cube, and flipping one $S^z_\alpha$ results in flipping a small domain frame defect perpendicular to $\alpha=a,b,c$, as in Fig[\ref{frame}].
Thus, the ground state will look like a phase in which these defects have proliferated.
However, these flipping operations act with a phase that depends on the domain frame configuration on surrounding sites.  

\begin{figure}[h]
  \centering
      \includegraphics[width=0.35\textwidth]{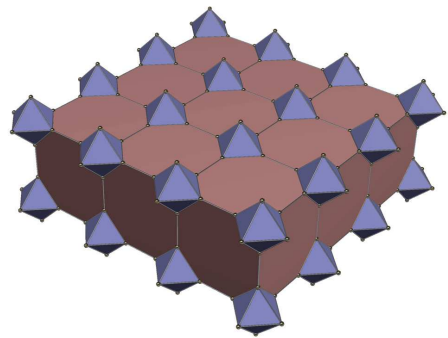}
  \caption{The dual lattice of our BCC plaquette Ising model composed of truncated cubes, each with an octahedron (with 3 perpendicular plaquette) intersecting every corner.} 
  \label{ccc}
\end{figure}

We now work on the dual lattice, where faces are replaced by links penetrating them, and volumes are replaced by vertices.  
This results in a simple cubic lattice, but with each vertex decorated as an octahedron, as shown in Fig~\ref{frame}.
The domain frame configuration is obtained by drawing a line penetrating the center of any square or triangular plaquette with $\prod S^z = -1$. 
The domain frames are therefore graphs on the edges of a dual lattice whose edges penetrate the centers of the square and triangular plaquettes involved in defining our Hamiltonian.  The resulting dual lattice is the truncated cubic lattice shown in Fig.~[\ref{ccc}], with three perpendicular plaquettes forming a octahedra intersecting each corner of the cubic lattice. 
Our original corner spins $S_0$ live in the centers of truncated cubes, and the three types of spin dipoles live in the center of the octahedra.

\begin{figure}[h]
  \centering
      \includegraphics[width=0.4\textwidth]{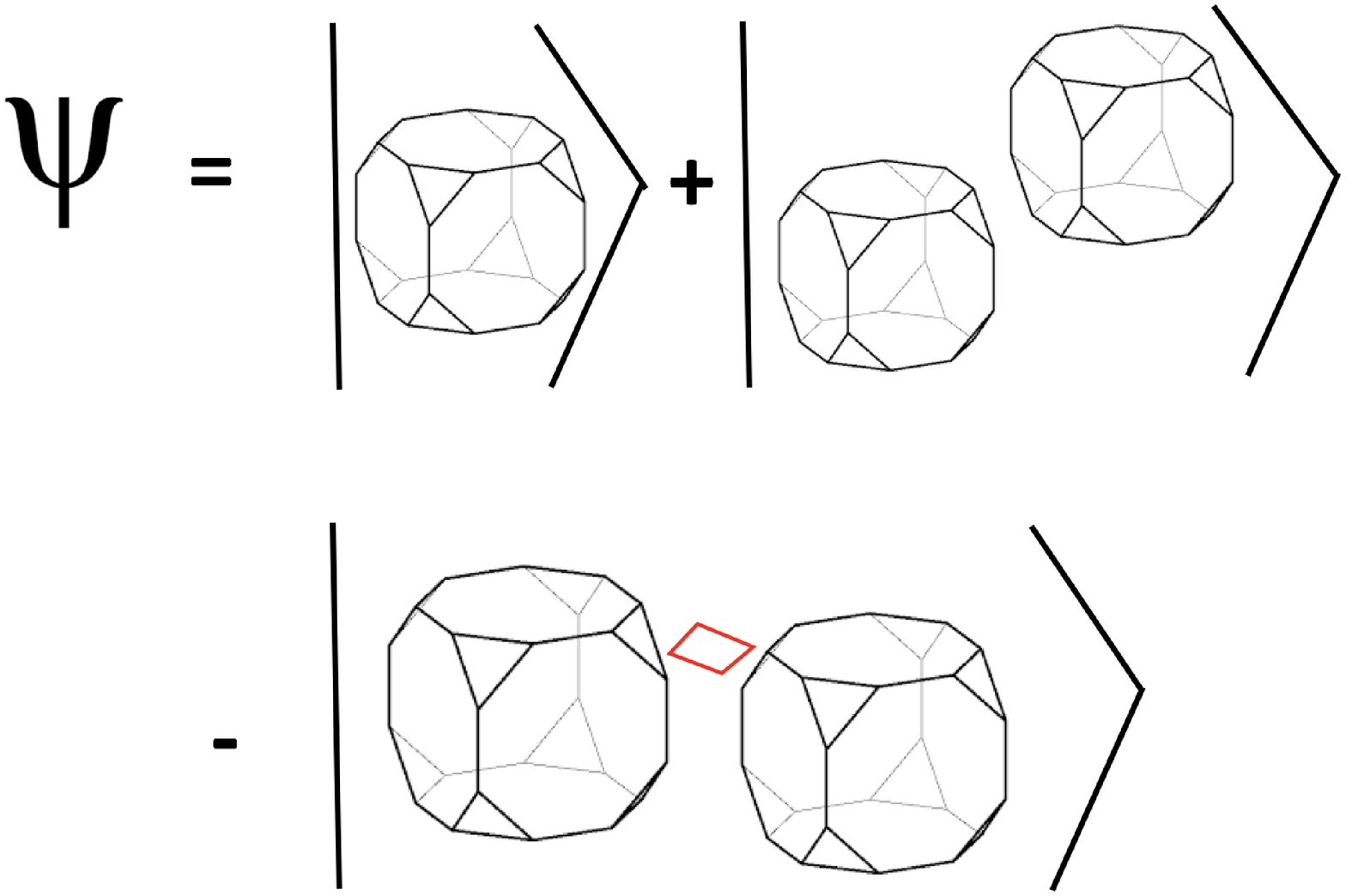}
  \caption{The paramagnetic ground state is a proliferation of all truncated cube and plaquette configurations. The coherent sum of such truncated cube and plaquette contains a special sign structure which counts the parity of domain frames in each configuration.} 
  \label{boxwave}
\end{figure}

The resulting ground state is depicted in Fig.~[\ref{boxwave}]: the coherent sum of all truncated cube (domain frame of $S^z_0$) and plaquette (domain frame of spin dipole $S^z_\alpha (\alpha \in a,b,c)$) configurations contain a special sign structure which counts the parity of \emph{loops} carried by truncated cubes and plaquette on all planes. 
To understand this sign structure consider a single planar layer of the truncated cubic dual lattice: within this 2D layer a given domain frame gives rise to a particular configuration of loops.  
The total number of loops is the sum of all the loops is all planes of all orientations.
Configurations with an odd (even) number of loops appears with a minus (plus) sign in the ground state. 
A truncated single cube, or a stack of several truncated cubes, creates an even number of surfaces,  resulting in an even number of loops. 
Hence, adding a truncated cube does not change the sign structure. 
Meanwhile, an isolated plaquette (domain frame for the spin dipole) creates an additional loop so adding these to the ground state generates a global minus sign.
As we will see in the next section, this sign structure will lead to a ``twisted" fracton order upon gauging the subsystem symmetry.

We can obtain a more generic (not exactly solvable) model for the topological subsystem paramagnet by adding to our Hamiltonian additional terms that respect the subsystem symmetry.  For example, we may add Ising plaquette interactions which energetically penalize the domain frames:
\begin{align} \label{Hfull}
& H=J H_t+h H_0  \nonumber\\
    &H_t=-\sum_{P} \prod_{i\in P}  S^z_{0,i}  -\sum_{\substack{T_\alpha,\\(ijk\in T_\alpha)}}  S^z_{0,i} S^z_{0,j} S^{z}_{\alpha,k}
\end{align}
The triangles $T_{\alpha=a,b,c}$ are defined as in Eq. (\ref{abc}) (see Fig.~[\ref{aaa}]).
Since $H_t$ respects all subsystem symmetries, provided that $h \gg J$, the model will remain in the paramagnetic phase.  On the other hand if $J \gg h$ the system will enter an ordered phase, where domain frames are confined.
We may also add transverse fields $S^x$, which when strong enough will drive the system into the trivial paramagnetic phase.


\subsection{Twisted fracton theory via gauging topological $Z_2$ plaquette paramagnet} \label{TwistedGaugingSec}

Next, we examine the effect of gauging this "twisted" topological plaquette paramagnet. We will show that the result is a phase with mutual lineon statistics. In the next subsection we use this result to argue the sysytem has symmetry protected gapless boundary modes.

To gauge the subsystem $Z^{sub}_2$ symmetry of the topological plaquette paramagnet in Eq.~[\ref{abc}], we proceed as described in Sec. \ref{Sec:GaugeReview}.  Because the rank-2 Ising gauge connection 
$\sigma^z$ lives at the centers of the square and triangular plaquettes  in our decorated cubic lattice,
%
 it is convenient to work on the dual lattice shown in Fig.~[\ref{ccc}].  Here the gauge connection $\sigma^z$ lives on edges,
and the original corner and dipole spins live in the centers of the truncated cubes and octahedra, respectively, of our dual lattice.  In this geometry,
the Gauss' law constraint  is:
\begin{align}  \label{GaussLaws}
&\prod_{i \in Cube } \sigma^x_i =S_0^x\nonumber\\
&\prod_{i \in P_\alpha } \sigma^x_i =S_\alpha^{x} 
\end{align}
The first product runs over all 36 links of the truncated cube surrounding the corner spin. 
The product in the next line over $P_\alpha$ for $\alpha=a,b,c$ runs over four links on {\it one} of the squares on the octahedron.  As an example, for $S^x_a$  we choose the square in the $b-c$ plane (and similarly for the remaining spin types). 
 
Note that the Gauss' law for the dipole spins $S_{a,b,c}^x$ is of a qualitatively different form, involving only fields along a single square, rather than along a surface surrounding the site.  Physically, this is because these spin dipoles carry vector charge, while spin $S_0$ carries scalar charge.

To obtain an effective Hamiltonian for the gauged theory, we begin by replacing all plaquette spin products with equivalent terms, minimally coupled to the gauge field:
\begin{align} 
& \prod_{i \in P} S^z_i  \rightarrow \prod_{i \in P} S^z_i \sigma^z_{P} \nonumber\\
\end{align}
Next, we replace all $S^x$ in Eq. [\ref{Hfull}] with appropriate products of $\sigma^x_P$, using the constraints [\ref{GaussLaws}]. Finally, if $J \ll h$, we keep only those products of terms proportional to $J$ which commute with the terms proportional to $h$.  The result is a commuting projector model describing the low-energy dynamics of the gauge theory deep in the paramagnetic phase: 
\begin{align} 
&H=-\sum_Y \prod_{i \in \text{Y}} \sigma^z_i - \sum_{Cube} (i)^{F_0} \prod_{i \in Cube } \sigma^x_i  - \sum_{P_\alpha}  (i)^{F_a} \prod_{i \in  P_a} \sigma^x_i   \\
&F_{\alpha\neq 0}=\sum_{i \in V_{P_\alpha}} (1-\sigma^z_i)/2 ,
F_0=\sum_{i \in V_{\text{Cube}}}  (1-\sigma^z_i)/2 
\label{gauge}
\end{align}
Here $Y$ is any set of three co-planar links entering a vertex, 
$V_{\text{Cube}}$ refers to the 48-links pointing outward from a truncated cube, and $V_{P_{\alpha}}$ ($\alpha=a,b,c$) refers to the 4-links in the plane perpendicular to $\alpha$ pointing outward from the octahedron.

In the $\sigma^z$ basis, the first term imposes a condition that the parity of the gauge flux entering any vertex is even.  
The remaining terms provide dynamics to the gauge field while preserving this parity at each vertex.  
However, these dynamics are qualitatively different from that of the $X$-cube model, since at low energies it effectively binds the charge ($\prod \sigma^x$) to the gauge flux (as measured by $F$).  

\begin{figure}[h]
  \centering
      \includegraphics[width=0.25\textwidth]{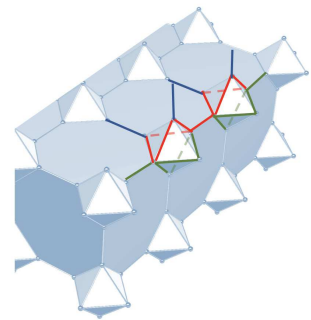}
  \caption{Line operator for creating a lineon charge. The flux excitation lives at the end of the line. The lineon is $1d$ particle, as the line operator cannot turn corners without creating additional excitations. } 
  \label{bbb}
\end{figure}

This charge-flux binding has important consequences for our model's fracton order.  The membrane operator that creates fractons is essentially the same as for the X-cube model, albeit modified to suit the different dual lattice geometry.  However the operator $L$ that creates pairs of lineons is qualitatively different.  
This acts along a ``decorated" line, as shown in Fig.~[\ref{bbb}], according to:
\begin{align} 
&L_{m}=\prod_{i \in red} \sigma^x_i \prod_{i \in blue  }  (i)^{\frac{1-\sigma^z_i}{2}}   \nonumber\\
&\prod_{i \in green } (i)^{\frac{1-\sigma^z_1}{2}+ \frac{1-\sigma^z_1\sigma^z_2}{2}+.... +\frac{1-\sigma^z_n \sigma^z_{n-1}}{2}+\frac{1- \sigma^z_{n}}{2}}  \ \ .
\label{lineon1}
\end{align}
The $\sigma^x$ operator along the red links flips the spin along the string, analogous to the lineon operator in the $X$-cube model.The terms acting on blue and green links are necessary to ensure that $L$ commutes with the Hamiltonian except near the endpoints of the line.

At each endpoint, we have $\prod_{i \in Y} \sigma^z_i = -1$ for both sets $Y$ at the vertex in question.  Note that as for the $X$-cube model, if the line along which the operator acts turns a corner, additional lineon excitations(i.e. vertices where prod $\sigma^z = -1$) are created.  
 As is the case for the Hamiltonian, though the operator contains factors of $i$, the matrix elements of $L$ are real provided that $\prod_{i \in Y} \sigma^z_i = +1$ -- i.e. provided that it does not cross paths with the {\it end-point} of another lineon operator. 
 At the end of the string, there is a $\pi$ gauge flux excitation, a lineon, which only moves along the string.
 
Apart from the lineon excitation in Eq[\ref{lineon1}], we have an anti-lineon excitation, created by the operator
\begin{align} 
&L_{\bar m}=\prod_{i \in red} \sigma^x_i \prod_{i \in blue  }  (i)^{\frac{1-\sigma^z_i}{2}}   \nonumber\\
&\prod_{i \in green } (-i)^{\frac{1-\sigma^z_1}{2}+ \frac{1-\sigma^z_1\sigma^z_2}{2}+.... +\frac{1-\sigma^z_n \sigma^z_{n-1}}{2}+\frac{1- \sigma^z_{n}}{2}}  \ \ .
\label{lineon2}
\end{align}
The lineon and anti-lineon pair fuses into to vacuum. Meanwhile, two lineons or anti-lineons fuse into a \emph{pair} of fractons at each end of the string,
\begin{align} 
&L_{e}=L_{\overline{m}}^2 = \prod_{i \in blue}  \sigma^z_i 
\end{align}

\subsubsection{Higher-rank Gauge theory}

Type-I fracton models, such as the X-cube model ( Eq.[\ref{HXCube}]), are generally described field theoretically by  discrete rank 2 gauge theories \cite{pretko2017generalized,ma2018fracton,Schmitz2017-ky,Pretko2017-nt,Vijay2016-dr,pretko2017subdimensional,pretko2017finite,bulmash2018higgs}. 
It is interesting to ask whether our gauged planar SSPT admits a similar description.  By analogy with exactly solvable lattice models in 2 dimensions\cite{levin2005string}, one might anticipate that the appropriate field theory contains a higher-rank discrete Chern-Simons term characterizing the nontrivial \emph{self}-statistics of the lineon excitation.  Here we outline what the lattice model suggests about this higher-rank Chern-Simons theory, whose physics we explore more thoroughly in Section \ref{cs}.

To proceed, we first review how the connection to higher-rank gauge theory arises in the X-cube model\cite{Vijay2016-dr,pretko2017generalized,pretko2017subdimensional,pretko2017fracton,ma2018higher,bulmash2018higgs}. We define a discrete rank two electric field  and gauge connection, $E_{ij}, A_{ij} $.  (In keeping with conventions for higher-rank gauge theories, here we will use subscripts $i,j,k$ as placeholders for the principal cubic axes, which we will denote by $x,y,z$).  These can be identified with the lattice plaquette spins (or dual lattice edge spins) $\sigma$ via
\begin{align} 
&\sigma^x= e^{i\pi E_{ij}},(E_{ij} \in {0,1})\nonumber\\
&\sigma^z=e^{i A_{ij}},(A_{ij} \in {0,\pi})\nonumber\\
&[A_{ij}(r),E_{ij}(r') ]=i \delta(r-r'),
\end{align}
Since $A_{ij}$ is associated with a plaquette in the $i-j$ plane, it is a symmetric tensor with only off-diagonal components. Alternatively, we can view $E_{ij}$ and $A_{ij}$ as defined on the dual lattice, in which case they reside on a link perpendicular to the $i-j$ plane.  For convenience, we will call the fields on link $\ell$ of the dual lattice $A_\ell, E_\ell$, while using the spatial indices $A_{ij}, E_{ij}$ to describe the continuum limit.

In terms of these discrete gauge fields, the constraint  in Eq.[\ref{GaugeConstr}] becomes
\be \label{GaussDiscrete}
\prod_{ \ell \in \partial C} e^{i \pi E_{\ell} } = e^{ i \pi \rho^f_C}
\ee
where $\ell $ runs over edges of the dual cube $C$, and we have defined a discrete fracton charge $\rho^f_C \in {0,1}$ residing at the center of the dual cube.  The cube term in Eq. (\ref{HXCube}) therefore enforces the condition that the ground state has no fracton charge.
The remaining vertex term in the Hamiltonian gives terms of the form
\be \label{BTerm}
 \text{exp} i \left[A_{xy}(\vec{r} ) + A_{xy}(\vec{r }- a \hat{z} ) + A_{yz}(\vec{r} ) + A_{yz}(\vec{r}- a \hat{x}  ) \right]
\ee

Eq. (\ref{GaussDiscrete}) can be viewed as a discrete lattice version of the usual Gauss' law for rank 2 scalar charge theories,
\begin{align} 
&\sum_{ij} \partial_{i} \partial_{j}  E_{ij}=p^{f} \ .
\end{align}
Meanwhile, Eq. (\ref{BTerm}) is simply the exponential of the $y$ component of the discrete magnetic flux 
\begin{align} 
&B_k=\epsilon^{ijk}\partial_{i} A_{jk}
\end{align}
about a given vertex.  
Combining the three vertex terms at a given vertex, and Taylor expanding in the lattice constant $A$, yields a term $ \sum_i (B_i)^2$ in the Hamiltonian.

As this suggests, one can check directly that the X-cube model describes a charge-2 Higgs phase of this rank-2 $U(1)$ gauge theory \cite{ma2018fracton,bulmash2018higgs}.  We note that  the identity $\sum_i B_i=\epsilon^{ijk} \partial_i A_{ik}=0$ arising from the continuum theory indicates that three orthogonal flux lines merging at a corner fuse into vacuum, and create no excitations.

For the twisted fracton model\cite{Ma2017-qq} in Eq.[\ref{gauge}], taking a product of the four co-planar $Y$-terms at a given vertex gives precisely Eq. (\ref{BTerm}), which again yields $\sum_i(B_j)^2$ upon taking the continuum limit. 
However, though the microscopic Gauss' law is the same, the terms involving $\sigma^x$ in Eq. (\ref{gauge}), which describe the constraint appropriate to the effective low-energy field theory of interest, contain a sign structure ($i^F$) which depends on the surrounding flux. 
For example, after projecting onto the low-energy Hilbert space the Gauss' law associated with a dipole spin $S_a = \text{exp}[ i \pi p^{dip}_z]$ in the center of a becomes:
\be
\prod_{\ell \in V_{P_z} } e^{i A_{\ell}/2} \prod_{\ell \in P_z} e^{i \pi E_{\ell}} = e^{i \pi p^{dip}_z}
\ee
and similarly for the other cubic axes.
Recall that $P_z$ is a square in the $x-y$ plane with edges along the $\hat{x} + \hat{y}$ and $\hat{x} -\hat{y}$ directions, while $V_{P_z}$ is the set of four edges emanating from the corners of this square, along the $\pm \hat{x}, \pm \hat{y}$ directions.

In finding the associated continuum constraint, we must specify a sign structure since in the discrete theory $E_{ij} \equiv - E_{ij}$.   As for the $X$-cube model, we interpret a product of $A_{ij}$ or $E_{ij}$ on spatially separated edges of the same orientation as {\it differences}, such that the leading-order terms in our continuum theory involve only derivatives of $A$ and $E$.  This choice ensures that our theory is gauge invariant.  Upon taking the continuum limit, it  leads a `modified `Gauss law': 
\begin{align} 
&\sum_i \partial_{i} E_{ik}=p^{dip}_k+\frac{B_k}{2\pi}\nonumber\\
&B_k=\epsilon^{ijk}\partial_{i} A_{jk} \ .
\label{att}
\end{align}
This can be viewed as forcing a $\pi$ gauge flux $B_z$ to the dipole charge $p^{dip}_z$. This reflects the fact that the lineon excitations that arise at the end of strings of with non-vanishing magnetic flux in the lattice model carries dipole charge.


The Gauss' law associated with a corner spin from Eq.~[\ref{att}] is
\be
\prod_{\ell \in V_{\text{C} }} e^{i A_{\ell}/2} \prod_{\ell \in \text{C}} e^{i \pi E_{\ell}} = e^{i \pi \rho^f_C}
\ee
where $C$ denotes a truncated cube, and $e^{i \pi \rho^f_C}$ describes the spin of the original lattice at the center of this cube.  Choosing an appropriate sign structure, and taking a continuum limit, we obtain a second Gauss' law:
\begin{align} \label{TwistGauss2}
&\sum_{ij} \partial_{i} \partial_{j}  E_{ij}=\rho^{f}+ \frac{1}{2 \pi} \sum_k\partial_k B_k
\end{align} 
where $\rho^f$ denotes a scalar charge associated with fracton excitations.  Again, this suggests a Chern-Simons like continuum theory, enforcing charge-flux binding.  

Notice that Eqs.~[\ref{att},\ref{TwistGauss2}] imply a relationship between the fracton and dipole charges in our model.  Specifically, if we take $\partial_k$ in Eq.~[\ref{att}]  and sum over the index $k$, we obtain Eq.~[\ref{TwistGauss2}], with 
\be \label{DipoleFractonRelation}
\rho^f = \sum_k \partial_k p^{dip}_k \ \ .
\ee
Thus, as for the X-cube model, a dipole can be viewed as a bound pair of fractons.


\subsection{Lineon boxing statistics and twisted fracton order} \label{BraidingSec}

 As the lineon string is decorated with polarized charge, two intersecting lineon operators anti-commute.  
 For example, if we have two  lineon operators $L_x,L_y$, acting on intersecting lines that run parallel to the $x$ and $y$ axes respectively, one can check that $L_x L_y=- L_yL_x$. This anti-commutation suggests the lineons have nontrivial (semion-like) statistics\cite{Ma2017-qq}. 
 
 
 To clarify the nature of this statistical interaction, we must first define the lineon analogue of the braiding and self-twisting operations that are used to diagnose statistical interactions for anyons in 2$D$. To do this, we must first understand how (if at all) the lineon operator can fluctuate without creating additional excitations.  In $2D$, anyons can be viewed as living at the ends of invisible, tensionless strings; since the strings can fluctuate and deform freely in space, only topologically non-trivial processes -- i.e.  braiding and twisting -- can give a universal result that characterizes the low-energy theory.  Thus far, we have described lineons as living at the ends of strings that {\it cannot} fluctuate, since adding corners to a string creates additional lineon excitations.  However this is not quite correct, though the string itself cannot fluctuate, it can be altered as follows: the line can turn if, in addition, another line emanates from the truncated corner.  This third line segment ensures that $\prod_{i \in ^Y} \sigma^z_i =+1$ for all edge triples at the given vertex, effectively by moving the lineon that would have been created by the corner to a different, more distant, location.  We can assemble several such corners into a truncated cube.  Isolated boxes of this type are simply truncated cube frames, identical to those generated by the Hamiltonian. 
 More generally they describe the 
 fluctuations allowed for our lineon operator.  


\begin{figure}[h]
  \centering
      \includegraphics[width=0.35\textwidth]{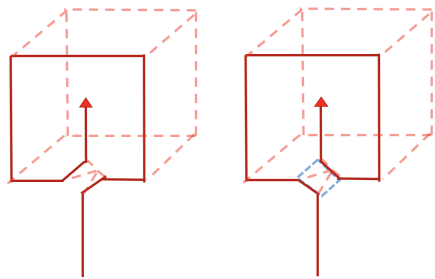}
  \caption{The trajectory of the self-rotation of lineon forms a domain frame where each corner of the frame contains three strings, one along each cubic axis. The left and right images depict the trajectory (red) of the self-rotation of the lineon by $\pm 2\pi$. 
The self-rotations of the lineon by $\pm 2\pi$ can be deformed into each other by adding a plaquette(blue) on the top layer of the cube.  This results in an overall phase, and hence semionic statistics.} 
  \label{box2}
\end{figure}

Our first example of how such boxing leads to non-trivial statistics is shown in Fig.~[\ref{box2}]. 
Due to the semionic nature of the lineon, the gauge flux contains a topological spin $\frac{1}{4}$ so that a self rotation of $4\pi$ accumulates a $\pi$ phase. 
The left (right) picture depicts the boxing process (red)  analogous to a self-twist of the lineon by $ 2\pi$ ($- 2 \pi)$: the lineon operator gets twisted, in addition to expanding into a box frame. 
In addition, these two opposite-chirality self-rotations can be deformed into each other by adding a plaquette (blue) on the front face. When our model is in its ground state, adding this extra plaquette results in a minus sign.  Thus if we act with a twisted lineon operator on this ground state, both positive and negative twists will be generated, {\it with opposite signs}.  As for the semion topological order in $2D$~\cite{Levin2012-dv}, it follows that the eigenstates of $\pm 2 \pi$ self-twisting have eigenvalue $\pm i$.   

\begin{figure}[h]
  \centering
      \includegraphics[width=0.3\textwidth]{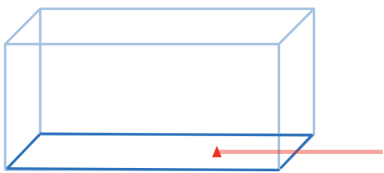}
  \caption{Braiding between two lineons. The blue cube frame is the trajectory of the lineon $B$.
  The string operator creating the lineon excitation $A$(red line) is in the same plane as the bottom surface of the cube. The blue line on the frame cross with the red string.
 } 
  \label{mutualflux}
\end{figure}

Our second example of a statistical boxing process is shown in Fig.~[\ref{mutualflux}].  We first act with a lineon operator running along the $\hat{x}$ direction (red line).  We then create a pair of lineons in the same plane, and use a boxing move to bring them around one end-point of this operator and re-annihilate them.  (The end result of this process is shown in blue).  Finally we re-annihilate the red lineon, returning the system to its ground state. (This last part holds because $L^2 = 1$ as long as no other lineon strings end along the line -- which we will assume to hold throughout this discussion).  The net process is described by
$L B L$, where $B$ is the boxing operator, which acts as the identity on the ground state.  
Provided that the blue lineon pair straddles the red lineon, as shown in the picture, the boxing operator crosses the lineon operator exactly once, and since perpendicular lineon operators anti-commute, we have $LBL = - B$.  Thus the ``box-braiding" of two lineons gives an analogue of semionic mutual statistics. 

Note that this lineon box braiding is different from the boxing operation described in Ref.~\cite{Ma2017-qq}, which describes a non-trivial statistical interaction between lineons and fractons.  In that case, the boxing operation gives a phase factor of $\pi$ whenever the fracton is inside the domain frame.  As Ref.~\cite{Ma2017-qq} noted, such a statistical interaction cannot occur between lineons, since the lineon can exit the domain frame through one of its faces without changing the phase.  In other words, lineons may have mutual statistics, but {\it only} with other lineons in the same plane so the trajectory of the box frame shall touch the other lineon string.

Finally, we would like to emphasize that the braiding procedure of lineons is distinct from the braiding of $2D$ particles or $3D$ loops. During the braiding procedure of $2D$ particles in the adiabatic limit, the system stays in the ground state manifold without any level crossing toward other excited states. However, in our boxing procedure for either self-rotation or mutual braiding for lineons, the initial and final states are in the ground state while the intermediate process contains additional excitations when the lineon changes its directions. 
Although these additional excitations are annihilated at the end, during the braiding procedure additional quasi-particles are created. 
As a result, this boxing procedure includes both a universal topological statistical phase and non-universal phase.

\subsection{Relation between lineon statistics and protected edge modes}

For symmetry protected topological phases protected by global unitary symmetries, it is generically true that after the symmetry is gauged, the flux excitations have nontrivial statistical interactions\cite{Levin2012-dv,chen2014symmetry,barkeshli2014symmetry,Kitaev2009-ju,ryu2015interacting,lu2016classification,bi2014anyon,thorngren2016gauging}. These nontrivial braiding statistics are intimately related to the existence of symmetry-protected gapless edge modes in the theory with global symmetry\cite{Levin2012-dv}. Here we give a similar argument to show that the nontrivial self-statistics of the lineons (which correspond to the fluxes of our rank 2 gauge field) ensure symmetry-protected gapless boundary modes.

\begin{figure}[h]
  \centering
      \includegraphics[width=0.25\textwidth]{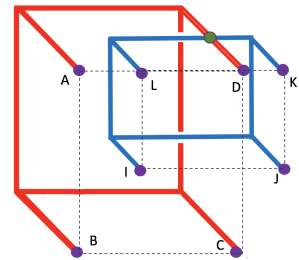}
  \caption{Adding two open domain frame operators to the ground state. The end points (black) of the frame lie on the surface. The two frames intersect at the green point in the bulk.} 
  \label{edge}
\end{figure}

Imagine we have a topological plaquette paramagnet with open surface, we can add two open box-frame to the ground state wavefunction
as in Fig[\ref{edge}], where the end point of the lineon string hits the surface. 

The wavefunction now becomes,
\begin{align} 
| \psi \rangle_1=W_r W_b | \psi \rangle_{gs}
\end{align}
$W_r,W_b$ refers to the blue/red domain frame operator in Fig[\ref{edge}].
These operators create domain frames in the bulk and let the lineon string end at the surface. As the bulk wave function still contains close frame configurations, such process does not create any excitation in the bulk. It merely creates a local defect on the edge where the plaquette interaction on the surface around a point has $\prod_{\in P} \sigma^z =-1$.
If the edge of the ground state is neither gapless nor breaks the $Z^{sub}_2$ symmetry, the local defect created by the end of the domain frame operator should be annihilated locally. The domain frame operator only changes the ground state within the region near the operator.
Hence, we can define an operator $U_a$ to annihilate the local defect at site $a$ on the edge. After such an operation, the state becomes our original ground state.
\begin{align} 
| \psi \rangle_{gs}=(U_A U_B U_C U_D W_r) (U_I U_J U_K U_L W_b) | \psi \rangle_{gs}
\end{align}

Now we choose another frame operation by switching the order of $W_r ,W_b$,
\begin{align} 
| \psi \rangle_2=W_b W_r | \psi \rangle_{gs}
\end{align}
Follow the similar process, we can annihilate the local defect on the edge and go back to the original ground state.
\begin{align} 
| \psi \rangle_{gs}= (U_I U_J U_K U_L W_b) (U_A U_B U_C U_D W_r)| \psi \rangle_{gs}
\end{align}
As $U$ is local, we can take the domain frame to be large enough so that each $U$ operator commutes with all others. As a result, we get $W_b W_r=W_b W_r$.
However, since there is an overlap between two string lines from the frame operator $W_b$,$W_r$ (depicted as the green point in Fig[\ref{edge}]), the nontrivial self statistics of the lineons bring about the anti-commutation relation $W_b W_r=- W_b W_r$. This contradicts our previous conclusion. Hence, our original assumption for a gapped symmetric edge fails. The edge must be either symmetry breaking or gapless, and the defect on the edge cannot be annihilate by local operators. 

The ungappable nature of our SSPT surface lies in the fact that the lineon excitation carries both flux and charge of the $Z_2$ symmetry.. For trivial plaquette Ising model in 3D, a subsystem symmetry invariant surface can be obtained via a lineon condensate on the surface. 
However, in the topological plaquette Ising model, the lineon carries subsystem $Z_2$ charge, and its proliferation would break the $Z_2^{sub}$ symmetry.

\section{Generalizations to larger symmetry groups: plaquette Ising Paramagnet with $Z^{sub}_2\times Z^{sub}_2$ symmetry} \label{tptp}

The subsystem $Z^{sub}_2$ SPT phase we developed in topological plaquette Ising model can be generalized to other discrete Abelian subsystem symmetries. 
In this section, we explore one such generalization, a plaquette Ising paramagnet protected by $Z^{sub}_2\times Z^{sub}_2$ symmetry.
Gauging the full subsystem symmetry yields a fracton order with two types of lineons and fractons -- one associated with each $Z_2$ subgroup.  A construction similar to that of the previous subsection can endow one or both of these lineons with semionic self-statistics.  However, in this case there is a new possibility: the two lineons can also  have nontrivial mutual statistics.  This leads to a $(Z_2)^3$ classification of the possible SPT phases, the details of which are described in Appendix \ref{tptpappen}.  Here we will focus on one example exhibiting this new type of mutual statistical interaction, in which the lineons have both semionic self-statistics, and non-trivial mutual statistics; the remaining cases are treated in the Appendix.

We will describe the Hamiltonian on the same truncated lattice as Fig[\ref{ccc}], with all degrees of freedom being doubled.  
There are two types of spin $s_0,q_0$ in the center of cube and two set of spin dipoles $(s_a,q_a),(s_b,q_b),(s_c,q_c)$ in the small octahedron. The Hamiltonian is 
\begin{widetext}
\begin{align} 
&H_0=- \sum_i \left ( s_{0,i}^x i^{F_0}-s^{x}_{a,i} i^{F_{a,i}}-s^{x}_{b,i} i^{F_{b,i}}-s^{x}_{c,i} i^{F_{c,i}} \right )- \sum_i \left ( q_{0,i}^x i^{F_0}-q^{x}_{a,i} i^{F_{a,i}}-q^{x}_{b,i} i^{F_{b,i}}-q^{x}_{c,i} i^{F_{c,i}} \right )\nonumber\\
&F_{\alpha\neq 0}=\sum_{P_\alpha}  (1-\prod_{i\in P_\alpha} s^z_{0,i} q^z_{0,i} )/2\nonumber\\
&F_0=\sum_{\substack{T_\alpha, \\(ijk \in T_\alpha)}}  (1-s^z_{0,i} s^z_{0,j} s^{z}_{\alpha,k}q^z_{0,i} q^z_{0,j} q^{z}_{\alpha,k} )/2  
\label{abc2}
\end{align}
\end{widetext}
This Hamiltonian is similar to two copies of Eq [\ref{abc}], except the sign structure of the transverse field terms involves both $q^z$ and $s^z$. Importantly, the sign structure respects separate subsystem $Z_2$ symmetry for $s$ and $q$ spins, and retains the feature that all terms commute, rendering the model exactly solvable. 

We may proceed as above, and gauge the full subsystem symmetry by coupling $q^z,s^z$ to a pair of $Z_2$ gauge connections $\sigma^z,\pi^z$. The corresponding gauge theory is,

\begin{align} 
&H=-\sum_P \prod_{i \in \text{Y}} \sigma^z_i  - \sum_{Cube} (i)^{F_0}\prod_{i \in Cube } \sigma^x_i  -  \sum_{P_\alpha} (i)^{F_\alpha} \prod_{i \in  P_\alpha} \sigma^x_i  \nonumber\\
&-\sum_P \prod_{i \in \text{Y}} \pi^z_i  - \sum_{Cube} (i)^{F_0}\prod_{i \in Cube } \pi^x_i  - \sum_{F_\alpha}  (i)^{F_a} \prod_{i \in  P_\alpha} \pi^x_i  \nonumber\\
&F_{\alpha\neq 0}=\sum_{i \in V_{P_\alpha}} (1-\sigma^z_i \pi^z_i)/2,  
F_0=\sum_{i \in V_{Cube}}  (1-\sigma^z_i\pi^z_i)/2 
\label{gauge2}
\end{align}
The ground states and low-lying excitations of this gauge theory can be studied by the same methods  employed for Eq.~[\ref{gauge}]. As promised, there are two types of fracton and lineon excitations. 
As in the previous section, the two types of lineons can be created using a decorated line operator of the type shown in Fig[\ref{bbb}].  In this case, we take
\begin{align}  \label{LaLb}
&L_A=\prod_{i \in red} \sigma^x_i \prod_{i \in blue  }  (i)^{\frac{1-\sigma^z_i \pi^z_i}{2}}   \nonumber\\
&\prod_{i \in green } (i)^{\frac{1-\sigma^z_1\pi^z_1}{2}+ \frac{1-\sigma^z_1\sigma^z_2 \pi^z_1\pi^z_2}{2}+.... +\frac{1-\sigma^z_n \sigma^z_{n-1}\pi^z_n \pi^z_{n-1}}{2}+\frac{1- \sigma^z_{n}\pi^z_{n}}{2}}\nonumber\\
&L_B=\prod_{i \in red} \pi^x_i \prod_{i \in blue  } (i)^{\frac{1-\sigma^z_i \pi^z_i}{2}}   \nonumber\\
&\prod_{i \in green } (i)^{\frac{1-\sigma^z_1\pi^z_1}{2}+ \frac{1-\sigma^z_1\sigma^z_2 \pi^z_1\pi^z_2}{2}+.... +\frac{1-\sigma^z_n \sigma^z_{n-1}\pi^z_n \pi^z_{n-1}}{2}+\frac{1- \sigma^z_{n}\pi^z_{n}}{2}}
\end{align}
As above, the decorated sign structure is necessary to ensure that these operators commute with the Hamiltonian away from their endpoints. For example, a $\pi_x$ string would create excitations along its entire length, rather than only at its endpoints. 
As in our previous model, there are also operators $\overline{L}_A, \overline{L}_B$ obtained from Eq. (\ref{LaLb}) by complex conjugation.  These satisfy $L_\alpha \overline{L}_\alpha = 1$, while the operator $L_\alpha^2 = \overline{L}_\alpha^2$ creates a pair of dipoles charged under both $Z_2$ symmetries.

The arguments of Sect. \ref{TwistedGaugingSec} can be used to argue that each type of lineon has semionic self-statistics in the sense discussed there.  
To see the `mutual statistics' between the two types of lineon, proceed as described in Sec. \ref{BraidingSec}, by using a boxing move to take a pair of B-type lineons around the endpoint of $L_A$, as shown in Fig.~[\ref{mutualflux}]. 
If the string operator $L_A$ creating the lineon excitation $A$ (red line in Fig.~[\ref{mutualflux}]) is in the same plane as the bottom surface of the cube, the bottom lines of the blue frame intersect with the red string during the braiding process. On such intersections the string operators $L_A,L_B$ anti-commutate.  Thus such a braiding process generates a $\pi$ statistical phase. This indicates the two lineons are mutual semions, which implies that $A$-type lineons are charged under $Z_2^B$ and vice-versa.  This mutual statistic is unchanged upon taking $L_A$ to $\overline{L}_A$, and similarly for $L_B$.

We may now consider the implications of the lineon statistics for the system's boundaries.  In the present example, each type of lineon is a self-semion, implying that there are two flavors of boundary modes that cannot be gapped.  In addition, however, since each lineon is charged under {\it both} $Z_2$ symmetries, the braiding statistics argument in section \ref{tp} also implies that it is not possible to gap out only one of these boundary modes, since doing so would lead to a contradiction for {\it both} types of lineon fluxes.


Similarly, a model of this type in which the lineon self-statistics are trivial, but where the two lineons are mutual semions in the sense discussed here, will also have subsystem symmetry protected boundary modes.
This leads to the $(Z_2)^3$ classification discussed in Appendix \ref{tptpappen}, with eight distinct choices of sign structure for the transverse field terms, each of which  leads to a distinct twisted fracton theory after gauging. 


\section{Continuum field theory of SSPT phases and SSPT phases with continuous symmetry} \label{cs}

As discussed in Sec.\ref{TwistedGaugingSec}, our lattice models for twisted fracton orders suggest that there exists a higher-rank version of Chern-Simons theory, in which magnetic flux is bound to dipolar charge via constraints of the form (\ref{att}, \ref{TwistGauss2}).  We now investigate the nature of this field theory in more detail.  

Specifically, we first write down a continuum Lagrangian for a rank-2 U(1) gauge field $A_{ij} = A_{ji}, A_{ii} =0$, and show that (1) it enforces the constraint (\ref{TwistGauss2}); (2) it is gauge invariant up to a boundary term; and (3) in the presence of a boundary, gauge invariance requires the addition of gapless boundary modes.  We will also briefly examine the associated statistics, which indicate non-trivial lineon braiding --  though we defer a careful discussion of quantizing this theory to future work.  

The rank-2 U(1) Chern-Simons theory is chiral: its boundary modes propagate in a chiral manner, and it has only a single type of lineon braiding statistic.  Thus this cannot be the field theory that describes our lattice models -- instead, it describes a 3D cousin of the integer quantum Hall effect.  Though our treatment shows only that the chiral gapless boundary modes are protected by $U(1)$ subsystem symmetry, we expect that, like their integer quantum Hall cousins and related systems with chiral dipole currents\cite{you2016response,gromov2017fractional}, these phases have an analogue of the thermal Hall  response\cite{haldane2009hall,cortijo2015elastic,cho2014geometry,hoyos2012hall,read2011hall,bradlyn2012kubo,read2009non}, such that the chiral boundary modes are robust even in the absence of $U(1)$ symmetry.  This suggests an integer classification for these higher-rank integer quantum Hall phases.


The SSPT ground states of Sec. \ref{TwistedGaugingSec}, in contrast, 
 have counter-propagating pairs of edge modes along the two cubic axes at each surface, and after gauging, the twisted fracton order has both ``semion" and ``anti-semion" -type lineons.  We argue that this   is described by two copies of this theory, with opposite chiralities.   We also introduce a higher-rank mutual Chern-Simons theory, which describe the models discussed in Sec. \ref{tptp}.



\subsection{U(1) subsystem-symmetric lattice models and coupling to higher rank gauge fields}

Before discussing the higher rank U(1) Chern-Simons theory, it is useful to briefly describe how U(1) subsystem symmetry arises at the lattice level.  This will give a clear picture of the nature of the dipole and fracton currents\cite{bulmash2018higgs,Ma2017-cb,pretko2017subdimensional,pretko2017fracton}, and show why the relationship (\ref{DipoleFractonRelation}) is natural in this context.  

\begin{figure}[h]
  \centering
      \includegraphics[width=0.35\textwidth]{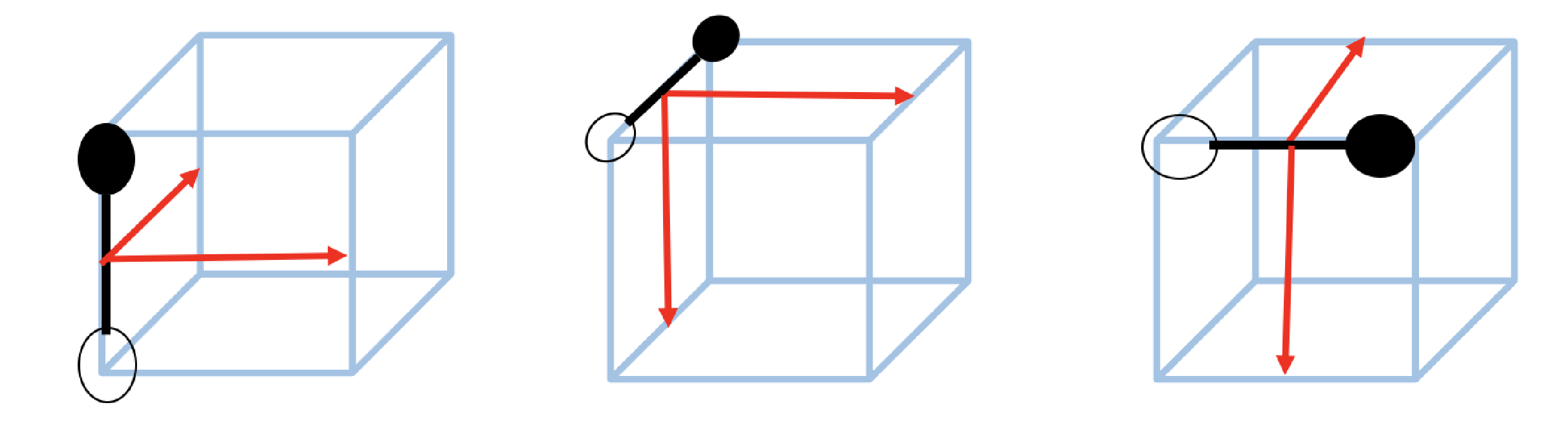}
  \caption{Fracton Hubbard model on the cubic lattice. Individual bosons have no mobility. However, a lattice-scale dipole comprised of a boson-hole pair can move in the plane perpendicular to the dipole moment.} 
  \label{fractondual1}
\end{figure}

We consider interacting bosons on a cubic lattice. Since subsystem symmetry is incompatible with boson hopping, the dominant symmetry-allowed kinetic term involves hopping a boson-antiboson pair (which can be interpreted as a lattice-scale dipole), as shown in Fig.~\ref{fractondual1}.   These dipole hopping moves preserve the net U(1) charge in each lattice plane, and the system has a subsystem $U^{sub}(1)$ charge conservation symmetry.   The dominant on-site interaction term compatible with subsystem symmetry is the usual Hubbard U.

In the continuum limit, the dipole hopping leads to the effective Lagrangian:
\begin{align} 
&\mathcal{L}=(\partial_t \phi)^2-t[(\partial_x \partial_y \phi)^2+(\partial_x \partial_z \phi)^2+(\partial_x \partial_y \phi)^2]\nonumber\\
&-U[\hat{n}-n_0]^2+...
\label{fourboson}
\end{align}
where $\hat{n}, \phi$ are the number and phase variables, respectively, that describe our bosons in the quantum rotor representation.  
A U(1) phase rotation in (for example) the plane $z= z_0$ leaves $\hat{n}$ invariant, and takes
\begin{align} 
&\phi(x,y, z=z_0) \rightarrow  \phi(z,y,z=z_0)- \alpha \ .
\end{align} 
When $U$ is large, the system is in a gapped Mott phase where 
the U(1) subsystem symmetry is unbroken.  

Let us now couple our bosons to a rank-2 U(1) gauge field.  In this case, the local gauge transformations take the form:
\begin{align} 
&\phi (r) \rightarrow   \phi(r)+ \alpha \nonumber\\
& A_{ij} \rightarrow  A_{ij}+\partial_{i}\partial_{j} \alpha \nonumber\\ 
&A_0 \rightarrow  A_0+\partial_0 \alpha
\label{gt}
\end{align}
The appropriate gauge-invariant kinetic terms are obtained by taking
\begin{align}
\partial_i \partial_j \phi \rightarrow \partial_i \partial_j \phi -  A_{ij} \nonumber \\
\partial_0\phi  \rightarrow \partial_0 \phi -  A_0  \ \ .
\end{align}
As usual, we see that $A_{ij}$ should be a symmetric tensor with only off-diagonal components. 
This leads to the 2-current:
\begin{align} 
&J_{0}= \partial_t \phi \nonumber\\
&J_{ij}= \partial_i \partial_j \phi \ \ .
\end{align}
Here $\partial_t \phi$ should be interpreted as the fracton charge density, while $\partial_j \phi$ can be viewed as a (lattice-scale) dipole, such that $J_{ij}$ represents a current in which a dipole oriented along the $\hat{j}$ direction propagates in the $\hat{i}$ direction.
Gauge invariance requires that this current satisfy the conservation law 
\be
\partial_t J_0 = \partial_i \partial_j J_{ij}
\ee


\subsection{Higher rank U(1) Chern-Simons theory} \label{CSSec}

Having clarified the origin of U(1) subsystem symmetry, and nature of the 2-current in our model, we now return to the question of higher-rank Chern-Simons theory.  
Ideally, we would be able to show directly that there exists a theory with U(1) subsystem symmetry where the underlying bosons (or fermions) can be integrated out in the presence of external gauge fields to yield our effective higher-rank Chern-Simons description, thereby establishing that it describes the rank-2 electromagnetic response of the model in question.  At present, however, we will simply write down a field theory and study the associated response, deferring a more detailed understanding of how it can emerge from models of the type described above for future work.  Indeed, we will see that the theory we write down here most likely represents the response of a fermionic system with U(1) subsystem symmetry.

Motivated by the constraints (\ref{att}, \ref{TwistGauss2}), we consider the Lagrangian
\begin{align} 
\label{LCS}
\mathcal{L} =& -\frac{1}{4 \pi } \epsilon^{ijk} (A_{jk}  \partial_t A_{ij} + 2 \partial_j A_0  \partial_i A_{jk } ) 
 - A_0 J_0 + A_{ij} J_{ij} 
\end{align}
Since $A_0$ is a Lagrange multiplier, it enforces the constraint
\begin{align} \label{EOM}
\frac{1}{2 \pi} \partial_i B_i = J_0 
\end{align}
Identifying $J_0\equiv \rho^f$, we have essentially recovered the modified Gauss' law of Eq. (\ref{TwistGauss2}), in the absence of electric fields. 

To see why this Lagrangian leads to gapless boundary modes, observe that under gauge transformations, 
\begin{align}
-\delta_{\alpha} \mac{L} 
=& \frac{1}{4 \pi} \epsilon^{ijk} ( \partial_j \partial_k \alpha \partial_t A_{ij} + A_{ik} \partial_t  \partial_i \partial_j \alpha  )\n
& + \frac{1}{4 \pi} \epsilon^{ijk} \partial_j \partial_k \alpha \ \partial_t \partial_i \partial_j  \alpha
\end{align}
where we have used the fact that $  \epsilon^{ijk}  \partial_i A_{jk } = B_k$ is explicitly gauge invariant.  

As for Chern-Simons theory, the action is gauge invariant {\it up to a boundary term}, implying that additional gapless degrees of freedom are required at the boundary to retain gauge invariance.  To be concrete, let's suppose that we work on a lattice with a spatial boundary, but with boundary terms such that all fields vanish as $t \rightarrow \pm \infty$.  Then we may discard boundary terms resulting from integrating by parts in time, which leaves us with
\begin{align}
\delta S
=& -\int_{ \partial_M}  \frac{1}{4 \pi} \epsilon^{ijk} \hat{n}_k (   A_{ij }  \partial_j \partial_t \alpha+  \partial_j  \alpha \partial_t \partial_i \partial_j  \alpha )
\end{align}
where $\hat{n}_k$ is the unit normal to the boundary.  From the first term, we deduce that to cancel the potential gauge anomaly, there must be an additional scalar field $\phi$ at the boundary, which transforms as $\phi \rightarrow \phi + \alpha$ under gauge transformations.  With this scalar field, the boundary Lagrangian becomes:
\begin{align} \label{Eq:Lbdy}
\mac{L}_{\text{Bdy}} =& -\frac{1}{4 \pi} \epsilon^{ijk} \hat{n}_k \left [ ( A_{ij} - \partial_i \partial_j ) \phi \right ]  \partial_j \left[ ( A_0 + \partial_t ) \phi\right] \n
& -
v^2 \left[ ( A_{ij } - \partial_i \partial_j ) \phi \right ]^2 
\end{align}
where we have added a gauge invariant kinetic term in the last line.  The resulting theory is explicitly gauge invariant.  

To understand what Eq.~(\ref{Eq:Lbdy}) means for the boundary, we set all gauge fields to 0, and  consider the scalar field with a boundary Lagrangian
\be
\mac{L} = -\frac{1}{4 \pi} \epsilon^{ij} \partial_i  ( \partial_j  \phi ) \partial_t  ( \partial_j  \phi ) -v^2\left[\epsilon^{ij} \partial_i \partial_j \phi \right]^2
\ee
Let us define
\be
\chi_i = \partial_i \phi
\ee
which is exactly the dipole charge along the $i^{th}$ direction.  
In terms of the $\chi$ fields, 
\be
\mac{L} = -\frac{1}{4 \pi}\left [  \partial_x  \chi_y \partial_t  \chi_y - \partial_y  \chi_x \partial_t  \chi_x \right ] - \frac{v^2}{2} \left [  (\partial_x \chi_y )^2  + ( \partial_y \chi_x )^2 \right ]
\ee
where we have used
\be
\partial_x \partial_y \phi = \partial_x \chi_y = \partial_y \chi_x
\ee
to write this in the most symmetrical way.  
This describes two chiral dipole currents -- a $y$-oriented dipole propagating along the $+\hat{x}$ direction, and an $x$-oriented dipole propagating along the $-\hat{y}$ direction -- at the boundary.  Note that as anticipated from our lattice description, the two dipole currents arise from the same underlying scalar field.  

\begin{figure}[h]
  \centering
      \includegraphics[width=0.35\textwidth]{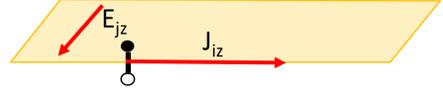}
  \caption{Higher rank Chern-Simons response for dipole current. By applying an electric field along $\hat{x} + \hat{y}$ on the $z=0$ plane, the z-oriented dipole moves along the $\hat{x} - \hat{y}$ direction on the $z=0$ plane.} 
  \label{fractondual2}
\end{figure}

It is natural to ask whether these gapless chiral boundary modes are associated with some form of bulk Hall-like response.  
Identifying $E_{ij} = \partial_t A_{ij} -  \partial_i \partial_j A_0$, the equations of motion derived from Eq. (\ref{LCS}) gives the following higher-rank electromagnetic response:
\begin{align} \label{HallCond}
\frac{1}{2 \pi}  E_{ij} = \epsilon^{ijk} J_{jk}
\end{align}
To see how this is related to the usual 2-dimensional Hall response, consider 
applying an external electric field $E_{iz}$ confined to the $z=0$ plane:
\begin{align} 
& E_{iz}=\delta(z) \partial_t  f(x,y) \ , \ \ E_{xy} = 0 \nonumber\\
&J_{yz}=\frac{1}{2\pi}E_{xz} \ , \ \  
J_{xz}= - \frac{1}{2\pi}E_{yz} \n
& J_{xy} = \frac{1}{2\pi} \left( E_{yz} - E_{xz}\right ) = 0
\end{align}
Interpreting $J_{jz}$ as a dipole oriented in $\hat{z}$ direction moving along $j$, we see that this is very much analogous to applying an electric field in the $\hat{x} + \hat{y}$ direction, and obtaining a $\hat{z}$- dipole current along the $\hat{x} - \hat{y}$ direction, as shown in Fig. \ref{fractondual2}.  

A second interesting question is whether Eq. (\ref{LCS}) leads to a statistical interaction between higher-rank gauge fields, and if so, of what type.
To investigate this, we observe that the Lagrangian (\ref{LCS}) indicates that after quantization,
\begin{align} \label{CSCom}
 \left[ A_{xy}, \frac{k}{2 \pi} (A_{xz} - A_{yz}) \right ] = i \n
 \left[ A_{xz}, \frac{k}{2 \pi} (A_{yz} - A_{xy}) \right ] = i \n
 \left[ A_{yz}, \frac{k}{2 \pi} (A_{xy} - A_{xz}) \right ] = i \n
\end{align}
where we have replaced the coupling $\frac{1}{4 \pi}$ with $\frac{k}{4 \pi}$ in Eq. (\ref{LCS}).  

To understand the implications of these commutation relations for quasiparticles, consider a gauge field configuration
\be
A_{yz} = \delta(y) \delta(z) \Theta(y) \Theta(L-y) \ , \ \ A_{xz}= A_{xy} = 0 \ \ .
\ee
The corresponding magnetic field configuration 
\be
B_z = - B_y  = \delta(y) \delta(z) \left( \delta(y) - \delta(y-L) \right )
\ee
corresponds exactly to the type of pattern we expect for lineon-type excitations, which as described in Sec. \ref{TwistedGaugingSec}, reside at isolated points where the magnetic field is non-vanishing.  In other words, lineon excitations live at the end-points of line operators of the form
\be
\hat{L}_k = e^{i \epsilon^{ijk} \int A_{ij} dr_k }
\ee
Note that unlike the usual Wilson lines, the $\hat{L}_k$ are necessarily associated with a particular direction on the cubic lattice, and cannot be continuously deformed.  (Instead, the appropriate deformation involves fluctuating domain frames).  
After choosing a gauge where $A_{xy} =0$, the commutation relations (\ref{CSCom}) imply that  two intersecting line operators $\hat{L}_x$ and $\hat{L}_y$  satisfy
\be
\hat{L}_x \hat{L}_y = e^{ i \frac{2 \pi}{k}}\hat{L}_y \hat{L}_x 
\ee
By analogy with the commutators of Wilson lines in ordinary Chern-Simons theory, this implies that the lineons have fractional statistics, with an exchange phase of $\pi/k$.  For example, the semionic statistics of the lineons in Sec. \ref{TwistedGaugingSec} correspond to $k=2$.  

We note that in the case $k=1$ discussed above, we conclude that the lineons are {\it fermions}.  This suggests that Eq. (\ref{LCS}) is best viewed as a higher-rank response for an underlying fermionic theory.

\subsection{SSPT states and higher rank Chern-Simons theory}

Finally, let us return to the question of a continuum field theory capturing our SSPT and twisted fracton phases.
In light of our understanding of the connection between ordinary Chern-Simons theory and SPT phases in 2D\cite{lu2012theory}, a natural guess is that the counter-propagating protected boundary modes are described by a mutual Chern-Simons theory 
\begin{align} 
&\mathcal{L}= 
-J_0^a A_0 + J_{ij}^a A_{ij} - J_0^b B_0 + J_{ij}^b B_{ij} 
\nonumber\\
&-\frac{1}{4\pi}\epsilon^{ijk}(\dot{A}_{ij} B_{jk} +\dot{B}_{ij} A_{jk}
-2 B_0 \partial_j \partial_i A_{jk}-2A_0 \partial_j  \partial_i B_{jk}
) \ .
\label{mutualcs}
\end{align}

To obtain the twisted fracton theory of Sect. \ref{TwistedGaugingSec}, we 
define 
\begin{align}
&E_{ij} = \frac{1}{4 \pi} \epsilon^{ijk}  
(B_{jk}- B_{ik} )
\end{align}
which is canonically conjugate to $A_{ij}$:
\be
[A_{ij}, E_{ij}] = i \ \ .
\ee
In terms of these fields,  
the constraints 
\begin{align} 
&\partial_j \epsilon^{ijk} \partial_i A_{jk}=2\pi J^b_0,\nonumber\\
&\partial_j \epsilon^{ijk} \partial_i B_{jk}=2\pi J^a_0
\end{align}
can be expressed 
\begin{align}  \label{GaussIsh}
&\partial_{ij} E_{ij}+\frac{\partial_k \epsilon^{ijk} \partial_i A_{jk}}{2\pi} = J^1_0 \n
& ~\partial_{ij} E_{ij}-\frac{\partial_k \epsilon^{ijk} \partial_i A_{jk}}{2\pi} =J^2_0
\end{align}
where 
\begin{align}
    J^1 = \frac{1}{2} (J^a +J^b ) \n
    J^2 = \frac{1}{2} (J^a - J^b)
\end{align}
Eq. (\ref{GaussIsh}) is exactly the modified Gauss' law in Sec.~\ref{tp}.  Note that in fact we obtain two types of sources, for which lineons bind opposite fluxes; these are naturally associated with the two types of lineons in the twisted fracton theory.  This is consistent with the lattice action, since at the lattice level there is no distinction between positive and negative flux.  

To better understand the meaning of the currents $J^1,J^2$,  note that the transformation
\begin{align}
& A_{ij}^1=A_{ij}+B_{ij},~A_{ij}^2=A_{ij}-B_{ij}, \nonumber\\
&A_{0}^1=A_{0}+B_{0},~A_{0}^2=A_{0}-B_{0}
\end{align}
can be used to express the Lagrangian (\ref{mutualcs}) as two opposite-chirality copies of Eq. (\ref{LCS}):
\begin{align} 
&\mathcal{L}= -J^1_0 A^1_0 + J^1_{ij} A^1_{ij}-J^2_0 A^2_0 + J^2_{ij} A^2_{ij}\nonumber\\
&-\frac{1}{8\pi}\epsilon^{ijk}(A^1_{jk} \dot{A}^1_{ij} - A^2_{jk} \dot{A}^2_{ij} -2 A^1_0 \partial_j \partial_i A^1_{kj}+2A^2_0 \partial_j  \partial_i A^2_{jk})
\label{Doublecs}
\end{align}
 The factor of $1/2$ in the Chern-Simons coupling reflects the fact that if $A$ and $B$ are periodic modulo $2 \pi$, with charges quantized in integers, then $A^1$ and $A^2$ are periodic modulo $4 \pi$, with charges quantized in half-integers.  


The arguments in Sec.\ref{CSSec} simply imply that the boundary action (\ref{Doublecs}) describes counter-propagating chiral dipole fields $\partial_j\phi^1,\partial_j\phi^2$, which move along the $\hat{i}$ direction of an $i-j$ surface.
This is the natural higher-rank analogue of the boundary of the 2D $Z_2$ SPT phase.

\section{Symmetry enriched fracton phase in $3d$ topological plaquette paramagnetic phase with subsystem $\mathcal{T}^{sub}$ symmetry} \label{set}
Thus far, we have explored the relationship between twisted fracton order and subsystem SPT phases with unitary symmetry.  We have shown that gauging these SSPT phases leads to a twisted fracton order with nontrivial lineon braiding statistics.

We now turn our attention to the question of what happens if we gauge only part of the subsystem symmetry.  Specifically, we will consider symmetries of the form $G = Z_2 \times H$, and ask what happens when we gauge only the $Z_2$ part of the subsystem symmetry.  This will lead to a phase with the fracton order of the $X$-cube model, but where either fractons or lineons may carry fractional charges of $H$.  These phases are the fracton equivalent of symmetry-enriched topological (SET) phases\cite{lu2016classification,Gu2014-lj,else2014classifying,chen2015anomalous,mesaros2013classification,hung2013quantized,chen2016symmetry}. 

As a first example, consider gauging only one of the $Z_2$ symmetries in Section \ref{tptp}.  The resulting model has only one class of lineons, which do not have mutual statistics.  However, the lineon operators are decorated by phases that depend on the domain frame configuration of the remaining $Z_2$ subsystem symmetry.  The result is that these lineons carry a fractional (half-) charge under this ungauged $Z_2$, which can be measured by the same boxing operation used to detect the mutual lineon statistics in the gauged theory.

One obvious way to construct fracton SET phases is to use a coupled layer construction~\cite{Ma2017-qq}, where the layers being coupled are 2D SETs, rather than simply 2D topologically ordered phases, taking care to ensure that the coupling results in a model with subsystem symmetry.  In this construction lineons are bound states of flux excitations from two distinct intersecting 2D layers.  Suppose that the flux excitation in each layer is fractionally charged under $H$, and that the layers are coupled in such a way that the final subsystem symmetry is equivalent to acting with $H$ on a single $2D$ layer.  Then the lineon is clearly fractionally charged under the subsystem symmetry operations in either of the two layers associated with the bound fluxes of the original SET. Indeed, this is the nature of the half-charge carried by the lineons in our $H = Z_2$ example given above.

In light of this, it is particularly interesting to consider fracton SET phases where $H = \mathcal{T}$, for which the analogue of fractional charge is to carry the projective representation with $\mathcal{T}^2 = -1$.  This is because anti-unitary symmetries cannot truly be subsystem symmetries: once the different layers are coupled, complex conjugation must act globally, since it must conjugate the couplings between different planes.  (More explicitly, following  \cite{2018arXiv180302369Y}, we define a ``subsystem time reversal symmetry" $\mathcal{T}^{sub}$ as global complex conjugation, possibly combined with a subsystem spin rotation $i \sigma^y$.)  This opens the possibility of fracton SET phases that do not admit a coupled layer construction. 

To generate a $\mathcal{T}$-enriched fracton order, we draw inspiration from a well-known construction of 2D SPT phases with symmetry group $Z_2 \times H$, in which  the Ising domain walls are decorated with a $1D$ SPT with global symmetry $H$.  For example, if $H = \mathcal{T}$ is time reversal symmetry, we can decorate each Ising domain wall with an AKLT chain\cite{chen2014symmetry,you2016decorated,you2016stripe,you2014symmetry}. In the SPT, this means that domain walls ending on the boundary transform as Kramers doublets under $\mathcal{T}$.  Upon gauging the $Z_2$ symmetry we obtain an SET, whose flux excitations host Kramers doublets. 

Evidently, the notion of decorating domain walls cannot be directly imported into the context of subsystem symmetries where, as we have seen, the natural analogue of the domain wall is a domain frame.  Because the domain frames necessarily contain corners where three mutually orthogonal lines meet, we cannot simply decorate them with AKLT chains: one of the three chains would have to end at a corner, leaving a dangling spin 1/2 which cannot be gapped without breaking $ \mathcal{T}$ symmetry.  

Instead, to decorate our domain frames we introduce a new quasi-1 dimensional paramagnet which we call a ``valence tube solid" (VBT).  
This allows us to decorate domain frames such that corners are fully gapped, while domain frames ending on the system's boundary are bound to a Kramers doublet.  Gauging $Z_2^{sub}$ leads to a fracton phase with the same fracton order as the $X$-cube model, but where the lineons carry Kramers doublets.

\subsection{Valence tube solid}

\begin{figure}[h]
\centering
      \includegraphics[width=0.3\textwidth]{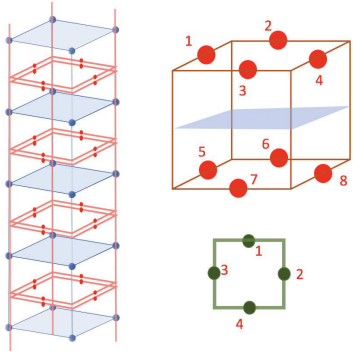}
\caption{Left: The valence tube solid on a vertical domain frame line. $\sigma$ spins live at the corners of the blue plaquettes.  Right: On the blue plaquette, $\sigma^z_a \sigma^z_b \sigma^z_c \sigma^z_d=-1$. The Hamiltonian then projects the red cube  into the  8-spin cluster entangled state $| \chi \rangle$. 
}
\label{twofig}
\end{figure}

Before introducing our Hamiltonian, let us briefly describe the valence tube solid.  
Consider a single cube with a spin-$1/2$ on each of the 8 edges in the $x-y$ plane, as shown in Fig.~\ref{twofig}.  We may project these 8 spins onto a Kramers singlet as follows~\cite{2018arXiv180302369Y}.  First, for a pair of spins (let us call them $s_1$ and $s_2$) separated in the $\hat{x} + \hat{y}$ direction, we project onto the two states:
\begin{align}
    |0\rangle_{12} = |\uparrow \rangle_1 |\downarrow \rangle_2 - |\downarrow \rangle_1 |\uparrow \rangle_2 \nonumber \\
    |1\rangle_{12} = |\uparrow \rangle_1 |\uparrow \rangle_2 + |\downarrow \rangle_1 |\downarrow \rangle_2 
\end{align}
This interaction does not preserve the full spin rotation invariance, but it does respect time reversal symmetry, which acts on each spins as $\mathcal{K} ( i \sigma^y)$.  Note that applying the spin rotation $i \sigma^y$ to both spins leaves the states invariant, while the states $|0\rangle_{12}, |1\rangle_{12} $ transform as a Kramers doublet under the action of $i s^y_1$ (or equivalently, of $i s^y_2$).
Next, on each plaquette perpendicular to $\hat{z}$, we project the four remaining states onto the two states: 
\begin{align}
    | \alpha\rangle_{1234} =  |0\rangle_{12}  |1\rangle_{34} - |1\rangle_{12}  |0\rangle_{34} \nonumber \\
    | \beta\rangle_{1234} =  |1\rangle_{12}  |1\rangle_{34} + |0\rangle_{12}  |0\rangle_{34}
\end{align}
Again this projection preserves time reversal symmetry.  Also, note that acting with $i s^y$ only on spins $1$ and $4$ (or equivalently, $2$ and $3$) takes $| \alpha\rangle_{1234} \rightarrow -| \beta\rangle_{1234}, | \beta\rangle_{1234} \rightarrow | \alpha\rangle_{1234} $.  
Finally, we couple the two 2-state systems on the cube's top and bottom faces, by projecting them onto a single state:
\begin{align}
    | \chi \rangle_{c}=\frac{1}{\sqrt{2}}(| \alpha \rangle_{1234}|\beta \rangle_{5678}-|\beta \rangle_{5678}| \alpha \rangle_{1234}) \ \ .
\end{align}
where $c$ denotes the cube, with the participating edge spins labelled $1$ through $8$, as shown in Fig. \ref{twofig}.
This picks out a unique ground state for the cube, which is invariant under acting with $\prod (i s^y)$ individually on the quadruples of spins lying in any one of the  $x-z$, $y-z$, or $x y$ planes.  We will call the Hamiltonian that selects this unique state the cube cluster interaction, given by:
\begin{align}
H_{CCI} = | \chi  \rangle_{c}  \langle \chi |_{c}
\end{align}

Next, imagine a tube containing a chain of such cubes aligned along $z$ direction, with two spin-$1/2$'s  on each edge in the $x-y$ plane, as shown in Fig.~\ref{twofig}.  On each edge one spin-$1/2$ participates in the cluster interaction of the cube above, and the other joins the cube below.  This gives a Hamiltonian of commuting cluster terms; its gapped symmetric ground state is our valence tube solid.  Note that our Hamiltonian leaves one plaquette at each end of the tube.  Applying the projector on this plaquette $P$ leaves a pair of states $|\alpha\rangle_P, |\beta \rangle_P$.


\begin{figure}[h]
\centering
     \includegraphics[width=0.4\textwidth]{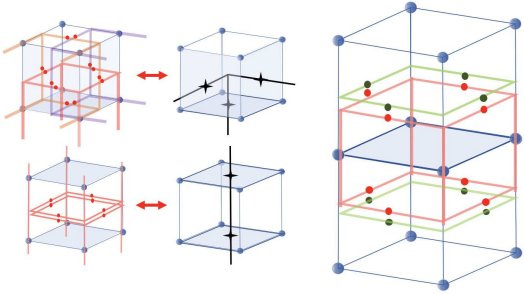}
     \caption{Left: When it comes to the corner of the frame, the three valence tube solids coming from the $x,y,z$ directions merge at the corner. Right: At the end of a frame line, there exist 4 free spins (green) which can be projected onto a 2-level system.}
\label{twofig2}
\end{figure}

The advantage of the valence tube solid construction is that there is a natural way to join three tubes in a corner without leaving unpaired spin-$1/2$'s. 
Consider a corner where three VBTs meet, as shown in Fig.~\ref{twofig2}.  
Notice that we choose the corner such that the tubes intersect at the mid-points of their edges, where the spins reside.  At each intersection there are two spin-$1/2$'s from each tube, for a total of 12 spins at the corner which participate in cluster interactions of cubes next to the corner. Thus the valence tube solid has a unique gapped ground state on frame structures with corners.  


 \subsection{Valence tube decorated fracton SET Hamiltonian with $\mathcal{T}^2 = -1$ lineons}

We now discuss how to use the VTS to decorate the $X$-cube model, obtaining a fracton SET in which the lineon transforms as a Kramers doublet under $T^{sub}$.  We will begin with a decorated version of the plaquette Ising model, which has $Z_2^{sub} \times \mathcal{T}^{sub}$ symmetry.  We will show that this model has the property that domain frames ending on the boundary carry $\mathcal{T}^{sub}$ Kramers doublets.  Following the procedure described in Sec. \ref{tp}, it is straightforward to gauge the $Z_2^{sub}$ symmetry and obtain the fracton SET. 

\begin{figure}[h]
  \centering
      \includegraphics[width=0.2\textwidth]{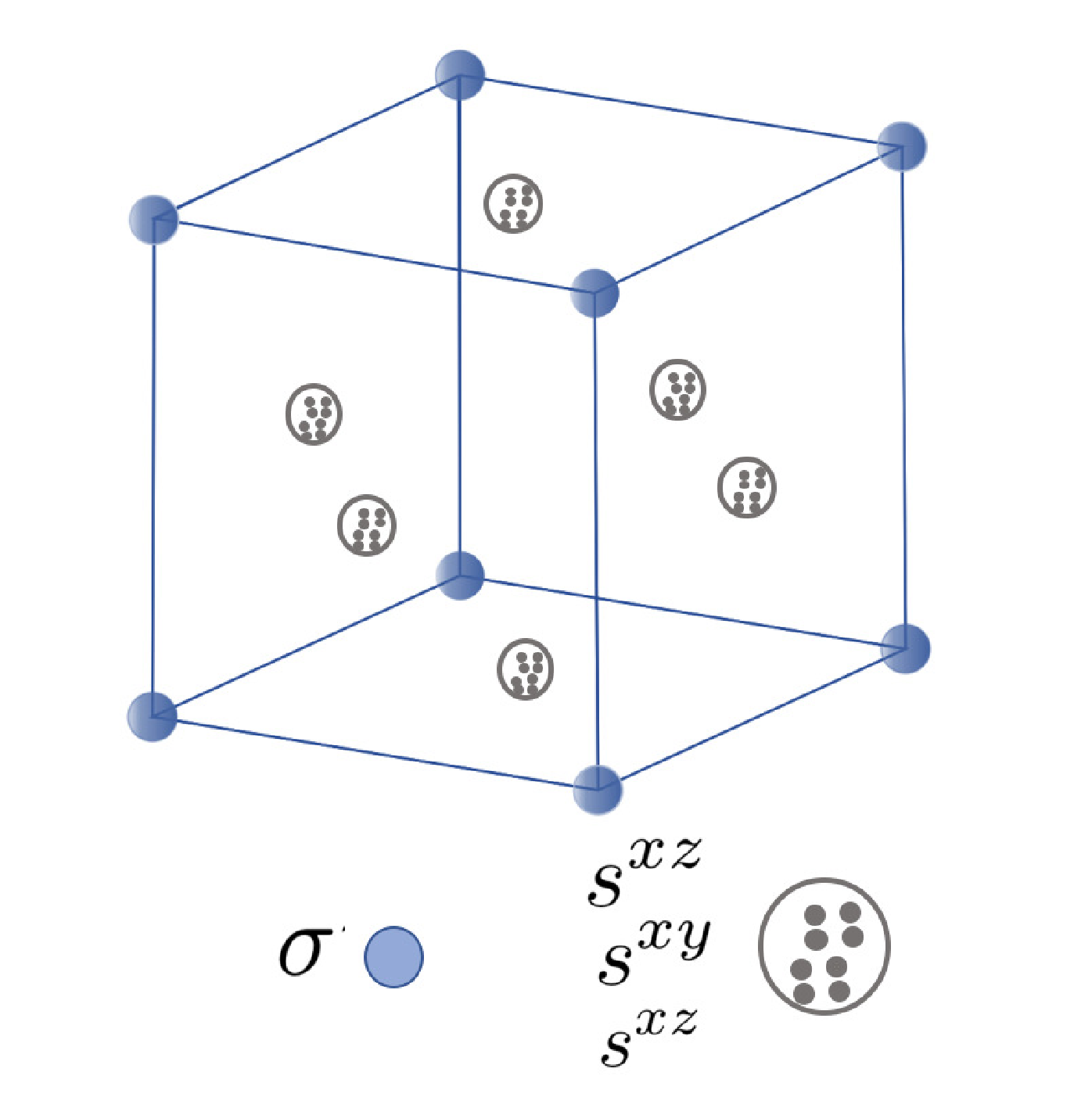}
  \caption{The blue site contains an Ising spin degree of freedom $\sigma$. At the center of each cubic face on the $i-j$ plane, there are 8 spin 1/2 degree of freedom depicted as the grey circles.} 
  \label{pc8}
\end{figure}

We start with the decorated cubic lattice shown in Fig.~[\ref{pc8}].  
Each blue site contains an Ising spin $\sigma$. At the center of each cubic face, there are 8 spin 1/2 degrees of freedom; we denote the spins in the $i-j$ plane by $s_{ij}$.  

The Hamiltonian consists of two terms:
\begin{align}
    H = H_{\text{PIM}} + H_{VBT}
\end{align}
The first term, which acts only on the $\sigma$ spins, is simply the plaquette Ising interaction and a transverse field: 
\begin{align} 
&H_{\text{PIM}}= \sum_{P_{ij}}  J \prod_{a \in P_{ij}}\sigma^z_a + h \sum_a \sigma^z_a \ \ .
\end{align}
The second term, $H_{VBT}$, is an interaction which effectively binds a VBT to each domain frame line of the $\sigma$ spins.
This can be done by introducing projection terms of the form
\begin{align} 
&H_{P_{xy}}=\left (1 -\prod_{i \in P_{xy}(R)} \sigma^z_i \right) | \chi\rangle_c \langle \chi|_c\nonumber\\
\label{p3}
\end{align}
where $c$ is a cube on the dual cubic lattice with $P_{xy}$ in its center (see Fig.~[\ref{twofig}]). 
 Thus spins assigned the edges of dual cubes that contain domain frame lines or corners form a VBT.
 $H_{\text{VBT}}$ also contains terms that project the remaining spins into on-site singlets, resulting in a unique symmetric ground state.


For example, consider a plaquette on the original lattice perpendicular to $z$-axis.  If $\prod_{i \in P} \sigma^z_i=-1$, the Hamiltonian projects the 8 spins above and below this plauqette into the state $\chi$(see Fig.~\ref{twofig}).  This leaves an even number of residual spin 1/2's on each side face, which are paired into on-site SU(2) singlets in the ground state. Likewise, the 8 free spins $s_{xy}$ on each $x-y$ face also form four on-site singlets (assuming that none of the neighbouring plaquettes contain domain frame lines).   

Deep in the paramagnetic phase, the resulting ground state wave function can be described as a uniform superposition of all domain frames, with each domain frame decorated by a valence tube solid.  
To understand the effect of this decoration, consider the effect of $\mathcal{T}^{sub}$ symmetry.  Here, we define $\mathcal{T}^{sub}$ symmetry as global complex conjugation, together with a rotation of all of the $s$ spins in a given plane:
\begin{align} 
&\mathcal{T}^{sub}=\mathcal{K} R_y^{sub} \nonumber\\
&R_y^{sub}: s_{ij}(r_k=k_0)  \rightarrow  i\tau_y ~s_{ij}(r_k=k_0),  \nonumber\\
&\mathcal{K} : i \rightarrow -i
\label{sym}
\end{align}
where $r_k$ is the $k$th component of the position vector $\vec{r}=(r_i,r_j,r_k)$, and $\tau_y$ is the $Y$ Pauli matrix acting on the spin-1/2 $s_{ij}$.  

Suppose that we act with $\mathcal{T}^{sub}$ on a plane covering the front face of the cube in Fig \ref{twofig}. The symmetry operator $\mathcal{T}^{sub}$ rotates spins 3 and 7 on the front face, and complex conjugates the wave function globally. Under such symmetry action, $|\chi\rangle$ is invariant. The plaquette entangled state $(|\alpha \rangle,|\beta \rangle)$ transforms as Kramers doublet.

\begin{align} 
&(i\tau^7_y i\tau^3_y) : |\alpha \rangle_{1234} \rightarrow |\beta \rangle_{1234}, |\beta \rangle_{1234} \rightarrow -|\alpha \rangle_{1234}, \nonumber\\
& |\alpha \rangle_{5678} \rightarrow |\beta \rangle_{5678}, |\beta \rangle_{5678} \rightarrow -|\alpha \rangle_{5678}, 
\end{align}

The Kramers doublet nature of the two plaquette states becomes significant when we consider a surface, where domain frame lines can terminate. Due to the valence tube decoration on the frame lines, the end point of the valence tube contains 4 free spins, one on each face parallel to the tube's direction (see Fig \ref{twofig2}). There is no way to fully gap out the four spins without breaking the subsystem $\mathcal{T}^{sub}$. The argument goes as follows: the $\mathcal{T}^{sub}$ is decomposed into a subsystem rotation operator $R_y^{sub}$ which rotates the spin on a plane, and a global complex conjugation. The $R_y^{sub}$ symmetry does not allow any spin interaction in the $s^x,s^z$ channel among the four spins, as they belong to four different planes. Hence, the only allowed interaction involves products of an even number of $s^y$ operators, which cannot  fully lift the degeneracy: since the four spins are in different planes, the ground states of such an interaction are Kramers singlets under some $\mathcal{T}^{\text{sub}}$ symmetries.  
At best, the degeneracy can be reduced, for example by the $s^y$ interactions mentioned above, or using the projection operator:
\begin{align} 
&P_{ijkl}=\frac{1}{2}(| \alpha  \rangle_{ijkl}  \langle \alpha |_{ijkl}+| \beta \rangle_{ijkl}  \langle \beta |_{ijkl})\nonumber\\
\label{p3}
\end{align}
This projects the 4 spins into a two-level subspace $|\alpha  \rangle_{ijkl},|  \beta   \rangle_{ijkl}$. As $ijkl$ lives on four side faces belonging to four different planes, if we acts $\mathcal{T}^{sub}$ covering any one of the side faces, the $(| \alpha  \rangle_{ijkl},|  \beta   \rangle_{ijkl})$ pair transforms as Kramers doublet. Hence, the end of the VTS we decorate contains a free spin 1/2 with projective representation under $\mathcal{T}^{sub}$.

As the domain frames proliferates in the bulk, the surface state involves a superposition of all point defects  $\sigma^z_a \sigma^z_b \sigma^z_c \sigma^z_d=-1$ arising from  domain frame lines ending at the surface. These fluctuating point defects on the surface carries a Kramers doublet, so the surface spectrum is guarantee to be gapless as long as $Z^{sub}_2 \times \mathcal{T}^{sub}$ is preserved. 


When we gauge the $Z^{sub}_2$ symmetry, the gauge theory is akin to the X-cube model with fracton and lineon excitations. The lineon excitation, located at the end point of an open domain frame, contains a Kramers doublet under $\mathcal{T}^{sub}$ due to the VTS decoration of the frame lines. The gauged X-cube model subsequently becomes a symmetry enriched fracton phase\cite{chen2014symmetry,Chen2013-gq}.

\section{Subdimensional spin liquid with fracton-like spinons}\label{FractSet}
The VTS decoration on domain frames creates a fracton topological order whose lineon carries a Kramers doublet. One can regard this Kramers doublet as implying that the lineon is a `spinon'.  Since the lineon is a $1d$ sub-dimensional particle, the spinon excitation, although not strictly confined, can only move along a certain direction.  

At this point, it is natural to ask whether it is possible to decorate the fracton excitation with a Kramers doublet. In this section, show that the answer is yes.  Specifically, we construct a fracton model whose fracton excitation is a spinon, in the sense that it harbors a Kramers degeneracy. These spinons, though technically deconfined, are not mobile quasiparticle excitations.

\begin{figure}[h]
\centering
     \includegraphics[width=0.45\textwidth]{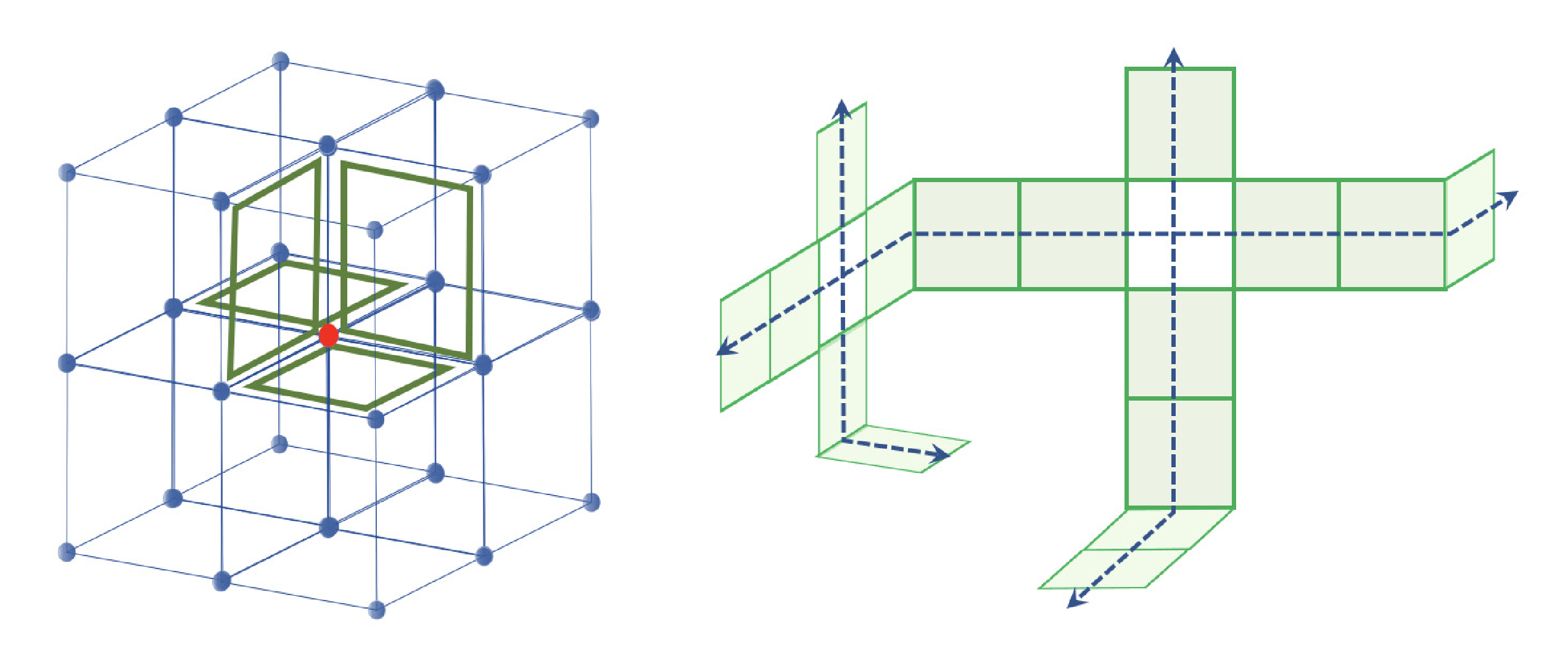}
     \caption{L: The twelve faces near a site must follow $\prod_{i \in Cube} \sigma^x_i=1$, so there can only appear even number of $\sigma^x_i=-1$ face near a site, each of which contains $| m \rangle_{ijkl}$ state on the plaquette. Here is one typical configuration near the site. R: One typical configuration for the plaquette singlet.}
\label{twofig3}
\end{figure}

Our construction begins with a cubic lattice, with twelve spins $S_{i_P}$ on each site $i$, and a single spin $\sigma_P$ at the center of each plaquette $P$, as shown in  Fig.~[\ref{twofig3}]. 
Each of the twelve spins $S_{i_P}$ is associated with one of the twelve plaquettes ($P=1 .. 12$) adjacent to the site $i$.
The full Hamiltonian is given by
\be
H = H_{\text{X-cube}} +\sum_P H_P + \sum_i H_i
\ee
The first term indicates that the $\sigma$ spins are coupled to each other via the $X$-cube Hamiltonian in Eq.~[\ref{HXCube}].  The second term is a coupling between each $\sigma$ spin and four $S$ spins: 
\be
H_{P} = -  \frac{1 - \sigma^x_P }{2} P^{(m)}_P 
\ee
Here $P^{(m)}_P = |m\rangle_P \langle m |_P$, where 
\begin{align} 
&| m \rangle_{P}\nonumber\\
&=\frac{1}{2\sqrt{2}}[(|0 \rangle_{i_P}|1 \rangle_{j_P}-|1 \rangle_{i_P}|0 \rangle_{j_P})(|0 \rangle_{k_P}|0 \rangle_{l_P}+|1 \rangle_{k_P}|1 \rangle_{l_p}) \nonumber\\
&-(|0 \rangle_{i_P}|0 \rangle_{j_P}+|1 \rangle_{i_P}|1 \rangle_{j_P})(|0 \rangle_{k_P}|1 \rangle_{l_P}-|1 \rangle_{k_P}|0 \rangle_{l_P})]
\end{align}
Here the indices $i_P,j_P,k_P,l_P$ indicate that at each of the four corners of the plaquette, we take the spin $S$ associated with the plaquette $P$.  

The final term is an on-site interaction acting on the remaining spins at each site:
\be
H_i = \sum_{ P,P'} \frac{1 + \sigma^x_P }{2}\frac{1 + \sigma^x_{P'} }{2} \vec{S}_{i_P} \cdot \vec{S}_{i_{P'} }
\ee
where the sum runs over all pairs of plaquettes $P,P'$ adjacent to the site $i$.  Let us denote by $Q$ those plaquettes on which $\sigma^x_Q = +1 $; then up to a constant $H_i = \frac{1}{2} \left ( \sum_Q S_{i_Q} \right)^2$.  The ground state of $H_i$ is thus the state for which these spins have the least total spin, which is $0$ for an even number, and $1/2$ for an odd number.

The resulting Hamiltonian has a unique gapped ground state and preserves $\mac{T}^{sub}$ symmetry.

When the $\sigma$ spins are in the X-cube ground state, the twelve faces near a site must follow $\prod_{i \in Cube} \sigma^x_i=1$.   In this case $Q$ is even, and $H_i$ selects a spin singlet ground state for the remaining site spins. 

\begin{figure}[h]
  \centering
      \includegraphics[width=0.25\textwidth]{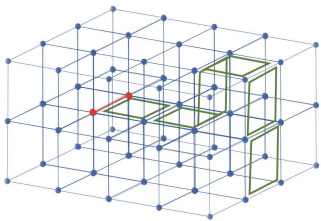}
  \caption{A pair of fracton excitations are shown in red. Each fracton carries a spin 1/2 Kramers doublet protected by the $\mathcal{T}^{\text{sub}}$ symmetry.  Here the tightly bound fracton forms a dipole, which can  move in a 2-d plane. The subsystem symmetry protects a Kramer's degeneracy even for the two spin-$1/2$'s on the dipole.} 
  \label{final}
\end{figure}

In the presence of a fracton excitation  (shown in Fig.~[\ref{final}]) at site $i$, $Q$ is odd, and the ground state of $H_i$ is doubly degenerate, with a total spin 1/2. In other words, each fracton carries a spin $\frac{1}{2}$ Kramers doublet. In addition,  a pair of fractons forming a dipole will  harbor a pair of spinons, as shown in Fig.~\ref{final}.  Because the two spins sit on different lattice sites, the resulting degeneracy cannot be completely lifted by any $\mathcal{T}^{sub}$ invariant interaction: the only interaction that respects this subsystem symmetry is of the form $S^y S^y$, which at best leaves a 2-fold degenerate ground state. This dipole is a $2d$ sub-dimensional particle, free to move in the plane perpendicular to its orientation. We have therefore constructed a  subdimensional spin liquid where the deconfined spinon excitation has no mobility, and a pair of (spatially separated) spinons are restricted to move on a $2d$ plane.

\section{Conclusion}

In the present work, we have introduced the notion of ``twisted" planar subsystem symmetry, and established its connection to both  subsystem symmetry protected topological (SSPT) phases, 
and twisted fracton orders. 
We have also introduced a higher-rank version of Chern Simons theory, which captures the physics both of the protected gapless boundary modes of the SSPT, and the nontrivial lineon statistics of the twisted fracton order.
Finally, we have 
discussed how subsystem symmetry can be used to {\it enrich} fracton orders, leading to subdimensional quasiparticles that exhibit symmetry fractionalization.  
This provides a foundation from which to explore new 3 dimensional phases with fracton order and/ or subsystem symmetry.

Numerous open questions about the relationship between subsystem symmetries and fracton orders remain.
(i) In 2D topological orders, the possible assignments of symmetry charges to anyons must satisfy several consistency conditions \cite{2002math......3060E,chen2015anomalous,barkeshli2014symmetry}, which are associated with the existence of 'tHooft anomalies \cite{qi2017folding,cheng2018microscopic,chen2016symmetry,wang2017anomaly}.  
Similar restrictions may exist for fracton theories, and it would be interesting to understand the underlying mathematical structure of these anomalies. (ii) In conventional 3D SPT phases, the gapless surface is necessary to cancel an anomaly of the global symmetry in the bulk theory.  This anomaly remains detectable even when the symmetry is broken at the surface, in the form of unusual edge states separating domains  where the symmetry is broken in distinct ways on the surface, such as the interface between domains with $\sigma_{xy} = \pm 1$ on the surface of a 3DTI.\cite{metlitski2013bosonic,chen2015anomalous}   In Sect. \ref{CSSec}, we have shown how the gapless boundary cancels a bulk anomaly in the $U(1) \times \overline{U(1)}$ SSPT. We expect that similar unusual boundary states exist at domain boundaries on symmetry-broken surfaces in this case. (iii) The anomaly realized on the surface of a 3D global SPT is closely related \cite{jian2018lieb} to obstructions to realizing a fully gapped Mott phase 
without breaking certain lattice symmetries, resulting in a generalization of the  Lieb-Schultz-Mattis theorem. We expect that this relation can be generalized to subsystem-symmetric phases, shedding light on the nature of  the Lieb-Schultz-Mattis theorem for these systems.
(iv) In 2D Chern-Simons theory, flux attachment can alter a particle's statistics from bosonic to fermionic. We anticipate that an analogue of this occurs for 3D fractonic matter, with a higher rank Chern-Simons term able to turn ``bosonic" lineons into fermionic ones.  By analogy with the half-filled Landau level, this could lead to a fermionic fracton phase,  with a fracton Fermi surface in the presence of a finite number of defects \cite{Prem2017-ql}.

{\bf Note added}

As we completed this work, we became aware of another paper \cite{2018arXiv180506899S} discussing twisted fracton orders, which has some overlap with our results.

\begin{acknowledgments}
We are grateful to Shinsei Ryu, Mike Pretko, Abhinav Prem for insightful comments and discussions. YY is supported by a PCTS Fellowship at Princeton University. FJB is grateful for the financial support of NSF-DMR 1352271 and the Sloan Foundation FG-2015-65927.

\end{acknowledgments}

\appendix 

\section{Classification of $Z^{sub}_2\times Z^{sub}_2$ SSPT via fracton gauge theory} \label{tptpappen}

In conventional SPT phases with Abelian symmetry, the classification for SPT states is equivalent to the identification of twisted gauge theories after we gauge the corresponding symmetry. If two short-ranged entangled states display the same gauge theory after symmetry gauging, the original states should belong to the same SPT class. In this appendix, we would apply such `SPT-gauge correspondence' to our SSPT with $Z^{sub}_2\times Z^{sub}_2$ symmetry. By figuring out eight distinct types of $Z_2\times Z_2$ fracton gauge theory, we conclude that our $Z^{sub}_2\times Z^{sub}_2$ SSPT has $(Z_2)^3$ classification. 

Take the topological plaquette Ising model in Section \ref{tptp}, 
\begin{align} 
H_0=&- \sum_i \left ( s_{0,i}^x i^{F_0}-s^{x}_{a,i} i^{F_{a}}-s^{x}_{b,i} i^{F_{b}}-s^{x}_{c,i} i^{F_{c}} \right ) \n
&- \sum_i \left ( q_{0,i}^x i^{F'_0}-q^{x}_{a,i} i^{F'_{a}}-q^{x}_{b,i} i^{F'_{b}}-q^{x}_{c,i} i^{F'_{c}} \right )
\end{align}
Here $F,F'$ is a general spin cluster operator which we would come about later. We can decorate the sign structure $F,F'$ in eight distinct ways.
\begin{widetext}
\begin{align} 
&\text{Type I}:F_0=1, F_\alpha=1;F'_0=1,
F'_\alpha=1\nonumber\\
&\text{Type II}:F_0=1, F_\alpha=1;F'_0=F'^{1}_0,
F'_\alpha=F'^{1}_\alpha\nonumber\\
&\text{Type III}:F_0=F^{1}_0, F_\alpha=F^{1}_\alpha;F'_0=1,
F'_\alpha=1\nonumber\\
&\text{Type IV}:F_0=F^{1}_0, F_\alpha=F^{1}_\alpha;F'_0=F'^{1}_0,
F'_\alpha=F'^{1}_\alpha\nonumber\\
&\text{Type V}:F_0=F^{2}_0, F_\alpha=F^{2}_\alpha;F'_0=F'^{2}_0,
F'_\alpha=F'^{2}_\alpha\nonumber\\
&\text{Type VI}:F_0=F^{2}_0, F_\alpha=F^{2}_\alpha;F'_0=F'^{2}_0+F'^{1}_0,
F'_\alpha=F'^{2}_\alpha+F'^{1}_\alpha\nonumber\\
&\text{Type VII}:F_0=F^{2}_0+F^{1}_0, F_\alpha=F^{2}_\alpha+F^{1}_\alpha;F'_0=F'^{2}_0,
F'_\alpha=F'^{2}_\alpha\nonumber\\
&\text{Type VIII}:F_0=F^{2}_0+F^{1}_0, F_\alpha=F^{2}_\alpha+F^{1}_\alpha;F'_0=F'^{2}_0+F'^{1}_0,
F'_\alpha=F'^{2}_\alpha+F'^{1}_\alpha\nonumber\\
&F^{1}_\alpha=\sum_{P_\alpha}  (1-\prod_{i \in P_\alpha}s^z_{0,i})/2,\nonumber\\
&F'^{1}_\alpha=\sum_{P_\alpha}  (1-\prod_{i \in P_\alpha}q^z_{0,i})/2,\nonumber\\
&F^{2}_\alpha=F'^{2}_\alpha=\sum_{P_\alpha}  (1-\prod_{i \in P_\alpha}s^z_{0,i} q^z_{0,i} )/2,\nonumber\\
&F^{2}_0=F'^{2}_0=\sum_{\substack{T_\alpha\\(ijk\in T_\alpha)}} (1-s^z_{0,i} s^z_{0,j} s^{z}_{\alpha,k} q^z_{0,i} q^z_{0,j} q^{z}_{\alpha,k} )/2 \nonumber\\
&F^{1}_0 =\sum_{\substack{T_\alpha\\(ijk\in T_\alpha)}} (1-s^z_{0,i} s^z_{0,j} s^{z}_{\alpha,k} )/2 \nonumber\\
&F'^{1}_0 =\sum_{\substack{T_\alpha\\(ijk\in T_\alpha)}} (1-q^z_{0,i} q^z_{0,j} q^{z}_{\alpha,k} )/2 \nonumber\\
\end{align}
\end{widetext}
All eight types of sign decoration gives the paramagnet model with exact solvability. After we gauge the $Z^{sub}_2\times Z^{sub}_2$ symmetry, one obtains eight fracton theory with distinct topological sectors. For lineon A and lineon B, each of them can have semion(boson) self-statistics. Meanwhile, the mutual statistical phase between lineon A and B can either be $\pi$ or $2\pi$. This creates 8 distinct fracton gauge theories which thereby identify 8 distinct SSPT phases (including the trivial phase as type I). For our model in Section \ref{tptp}, it belongs to type V where lineon A and B are both self-semion and mutual semion.

\end{document}